\documentclass{article}
\usepackage{amsmath, amssymb}
\usepackage[T1]{fontenc}
\usepackage[latin1]{inputenc}
\usepackage{graphicx}
\usepackage{caption}
\usepackage{bbold}
\usepackage{multirow}
\usepackage{graphicx}
\usepackage[para]{threeparttable}

\usepackage{arxiv}

\title{Improving Seasonal Forecast using Probabilistic Deep Learning}

\author{
Baoxiang Pan,
Gemma J. Anderson, 
Andr\'e Goncalves, 
Donald D. Lucas,
C\'eline J.W. Bonfils,
Jiwoo Lee
\\Lawrence Livermore National Laboratory, Livermore, CA 94550}

\begin{document}
\maketitle
\begin{abstract}
The path toward realizing the potential of seasonal forecasting and its socioeconomic benefits depends heavily on improving general circulation model based dynamical forecasting systems.  
To improve dynamical seasonal forecast, it is crucial to set up forecast benchmarks, and clarify forecast limitations posed by model
initialization errors, formulation deficiencies, and internal climate variability.
With huge cost in generating large forecast ensembles, and limited observations  for forecast verification,
the seasonal forecast benchmarking and diagnosing task proves challenging.
In this study, we develop a probabilistic deep neural network model, drawing on a wealth of existing climate simulations to enhance seasonal forecast capability and forecast diagnosis.
By leveraging complex physical relationships encoded in climate simulations, our probabilistic forecast model demonstrates favorable deterministic and probabilistic skill compared to state-of-the-art dynamical forecast systems in quasi-global seasonal forecast of precipitation and near-surface temperature. 
We apply this probabilistic forecast methodology to quantify the impacts of initialization errors and model formulation deficiencies in a dynamical seasonal forecasting system.
We introduce the \textit{saliency analysis} approach to efficiently identify the key predictors that influence seasonal variability. 
Furthermore, by explicitly modeling uncertainty using variational Bayes, we give a more definitive answer to how the El Ni\~{n}o/Southern Oscillation, the dominant mode of seasonal variability, 
modulates global seasonal predictability.
\end{abstract}


\section{Introduction}
\subsection{Dynamical seasonal forecast and its three barriers}
Global atmosphere-ocean-land coupled general circulation models (GCMs) that  simulate the dynamics and interactions of various climate subsystems are the workhorses for predictions on weather to climate scales. 
Over the past seven decades,
advances in computation, observation, modeling, and data assimilation have gradually extended the range of useful deterministic 
forecasts toward the predictability limit \cite{bauer2015quiet}.  
This limit of range, set by the chaotic nature of the atmosphere, is estimated to be around two weeks \cite{zhang2019predictability}.
Moving beyond this range, 
subseasonal to seasonal forecasts seek prediction sources from low-frequency climate signals, which give rise to predictability by constraining the climate variability through the forecasting period.

By better characterizing the key prediction sources and their impacts,  GCM-based seasonal forecasts have demonstrated markedly improved skill, 
offering crucial benefits to a wide range of societal sectors \cite{palmer2002economic}.
Despite the achievements, we highlight the following three barriers that hinder the improvement of dynamical seasonal forecast. 

The first barrier is model initialization.  Accurate initialization is a prerequisite for dynamical forecasts. However, existing initialization strategies often fail to constrain a model's state to faithfully match observations without introducing state \textit{shocks}   \cite{balmaseda2009impact, mulholland2015origin}. Furthermore, when compared to the atmosphere, the assimilation methods and observation networks for the ocean, cryosphere, and land are typically underdeveloped \cite{smith2012current}, inhibiting the full utilization of prediction sources from these climate subsystems.

The second barrier
is model formulation deficiencies. 
GCMs are simplified representations of the climate system. 
Their backbones are dynamical cores that compute the atmosphere and ocean dynamics in discrete geogrids. Besides the resolved dynamics, a suite of empirical parameterization schemes is used to account for the effects of unresolved  processes.
The considerable structural and parametric errors in parameterization can severely impair a model's forecast through multi-scale interactions. 

The third barrier concerns the
insufficient sampling of forecast spread. 
In dynamical forecasts,
uncertainties in initialization and model formulation can quickly grow and yield distinct, but possible results.
To estimate this forecast spread, we usually
create forecast ensembles by sampling plausible initial states and plausible model formulations.
A key question is: how many ensemble members are required to accurately infer the forecast spread?
There has been a growing recognition that initial states accommodating different climate signals can yield distinct forecast spreads,
resulting in variations of predictability \cite{pan2019precipitation, mariotti2020windows} and requirement on ensemble size.
Currently, we can not always afford large ensemble forecasts to accurately estimate forecast uncertainties. Also, the dependency of predictability on the initial climate state is not well-considered in deploying forecast ensembles or guiding forecast-related decision makings.

The difficulties in surmounting the outlined barriers 
have considerably slowed the pace toward more skillful seasonal forecasts. 
To accelerate forecast system development, we should set up reasonable forecast benchmarks \cite{best2015plumbing}, thereafter apply these benchmarks to clarify the limitations 
imposed by each aspect of the forecasting barriers \cite{smith2001disentangling}. 
Unfortunately, given the huge cost to initialize and run large-ensemble forecasts, and short history of climate observations 
for forecast verification, the seasonal forecast benchmarking and diagnosing task remains challenging. 

\subsection{Learning from climate simulations for seasonal forecast}
Two recent studies suggest that we can turn to the data-rich climate simulation ``model world'' to seek implications for seasonal forecasts \cite{ding2018skillful,ding2019diagnosing, ham2019deep}. For instance, Ding et al. \cite{ding2019diagnosing} drew analog from existing climate simulations that resemble observed ocean status as forecast ensembles, and found those analog forecasts match dynamical forecast systems in seasonal prediction for sea surface temperature and precipitation. 
This method could be interpreted as a \textit{k}-nearest neighbour approach \cite{altman1992introduction} from a machine learning viewpoint.
In the second study, 
Ham et al. \cite{ham2019deep} trained a deep neural network to predict the El Ni\~{n}o/Southern Oscillation (ENSO) in climate simulations, and found that, with moderate modifications and using relatively few observational records, the neural network model applies well in real-world forecast of ENSO, outperforming dynamical forecasting for lead time of up to one and a half years. 

The rationale behind these approaches is that, seasonal forecasts are conducted with identical or similar GCMs used in climate simulations. 
One can build analogical models using machine learning techniques to 
simulate the seasonal variability signal in climate simulations,
and thereafter apply this GCM-revealed relationship for practical forecast. 
While 
these attempts 
open the opportunity of
employing large-scale statistical models 
(i.e., deep neural networks) 
to learn from climate simulation big data for seasonal forecasts, 
we notice the following three gaps along this novel line of research. 
First, existing research focuses on limited regions or phenomena, and have not fully exploited the wealth of climate simulation information to improve seasonal forecast for a broad range of regions. 
Second, forecast uncertainty, which is indispensable in seasonal forecast and crucial in supporting forecast-related decision making, has not been rigorously represented. Finally, any solid progress toward improving seasonal forecast should contribute to tackling the three aforementioned forecast barriers noted in Sect. 1a.  
However,  
it is unclear how the analog seasonal forecast practices inform the identification or conquering of these forecast barriers.  

To address these  deficiencies,
we develop a probabilistic deep learning model
that leverages the
rich information from climate model simulations to enhance seasonal forecast capability and foster forecast diagnosis.
Specifically,
we summarize how key prediction
sources regulate seasonal variability in climate simulations by learning a generative model that samples possible 
seasonal-ahead climate status conditioned on the current state of the predictors.
By benchmarking dynamical seasonal forecasts with the proposed model, we 
demonstrate that current dynamical seasonal forecast systems have not fully exploited the forecast capability of existing GCMs. The impact of initialization error, GCM formulation deficiency, and internal climate variability are revealed.
The results help answer the following questions:
\begin{enumerate}
    \item Can we extend the data-driven seasonal forecasting paradigm 
    to quasi-global scale?
    \item How to represent uncertainty in a data-driven seasonal forecast? 
    \item How does this methodology inform potential improvement of dynamical seasonal forecasts? 
\end{enumerate}

We organize the rest of the paper as follows.
Section \ref{Section2} introduces the probabilistic modeling framework. Section \ref{Section3}  describes the data and experimental design. Section \ref{Section4}
compares our model with dynamical forecast systems in retrospective forecast experiments.
We discuss the implications for diagnosing and understanding dynamical seasonal forecast in Section \ref{Section5}. Conclusions are drawn in Section \ref{Section6}. 

\section{Methodology}
\label{Section2}

\subsection{Probabilistic modeling framework}

The model we develop here is a customized
Conditional Variational Auto-Encoder (CVAE) model (\cite{kingma2013auto,rezende2014stochastic,Kihyuk2015, ivanov2018variational}, \cite{kingma2019introduction}). It is a conditional generative model that marries probabilistic graphical models with deep learning.
Some of the key ideas here, such as variational inference and deep neural networks, may not be familiar to the present climate community, and relevant progress from the machine learning community is often built upon a series of recent work. Here, we 
provide a principle-based, step-by-step derivation of the probabilistic modeling framework. 
The specific model parametric form, architecture, and training strategy are thereafter explained. 

Given climate simulations generated by GCM ensembles, we sample snapshots of predictor $X$ and predictand $Y$ pairs from these simulations. 
For instance, $X$ could be the spatial distribution of upper ocean heat profile that has long memory and  extensive impact on seasonal variability \cite{ham2019deep, switanek2020present}, and $Y$ could be the corresponding spatial distribution of seasonal-ahead mean state of precipitation or near-surface temperature some months later.
Based on the extracted 
$(X,Y)$
pairs, we build a probabilistic forecast model by approximating the conditional probability distribution of $P(Y|X)$. 
This model has two sources of 
stochasticity. First, we focus on the impact of the considered predictor $X$, and discard explicit representation of  less-predictable processes,
such as the high frequency atmospheric component. Second, we inject model formulation noise by sampling climate simulation data generated by different GCMs.






We apply a parameterized function $p_{\theta}(Y|X)$ to approximate $P(Y|X)$. The learning process consist of optimizing ${\theta}$ such that $p_{\theta}(Y|X)$ is close to $P(Y|X)$ as revealed by our reference climate simulation data. While a suitable objective for optimizing $\theta$ is to maximize its conditional log likelihood
$\sum\log p_{\theta}(Y|X)$,
a direct optimization is difficult given the high dimensionality and potentially arbitrary dimension dependencies of $X$ and $Y$. 

To alleviate this difficulty, we
introduce a latent variable $Z$ to govern  the distribution of $Y$ conditioned on $X$.
This enables decomposing $p_{\theta}(Y|X)$ into simpler building-blocks of  $p_{\psi}(Y|X,Z)$ and $p(Z|X)$:

\begin{equation}
p_{\theta}(Y|X) = \int p_{\psi}(Y|X,Z)p(Z|X)dz
\label{eq:cond_distr_cvae}
\end{equation}
Here $p_{\psi}(Y|X,Z)$ is the conditional likelihood of $Y$ given $X$ and $Z$, with its parameter denoted by
$\psi$; $p(Z|X)$ is the conditional prior distribution of $Z$. 
We assume $p(Z|X)$ to be standard normal distributed, and drop its dependency on $X$. This assumption is acceptable with flexible enough $p_{\psi}(Y|X,Z)$, which allows mapping the stand normal distributed $Z$ to any preferred latent distribution form \cite{doersch2016tutorial}, meanwhile capturing the interaction between $X$ and $Z$ for inferring $Y$.
Once $p_{\psi}(Y|X,Z)$ is estimated,
we can generate samples of $Y$ by sequentially sampling $p(Z|X)$ and $p_{\psi}(Y|X,Z)$.


To apply maximum conditional likelihood estimation based on Equation \ref{eq:cond_distr_cvae}, we need to integrate over $Z$, 
which is computationally intractable. To tackle this problem, we resort to \textit{stochastic gradient variational inference} \cite{kingma2013auto, Kihyuk2015}.
Specifically, we narrow the integration space of $Z$ in Equation \ref{eq:cond_distr_cvae} by only considering
$Z$ values that are likely to generate $Y$ given $X$. Such information is given by the unknown posterior distribution of  $P(Z|X,Y)$.  In variational inference, we bring in an inference probability function $q_{\phi}(Z|X,Y)$ with parameter $\phi$ to approximate $P(Z|X,Y)$.
The error of this approximation is measured by the residual term as shown in the following equation (see appendix for derivation):

\begin{equation}
\log p_{\theta}(Y|X)=
\overbrace{\underbrace{E_{Z\sim q_{\phi}} [\log p_{\psi}(Y|X,Z)]}_{\text{Reconstruction term}}-\underbrace{KL\big(q_{\phi}(Z|X,Y)\Vert p(Z|X)\big)}_{\text{Regularization term}}}^{\text{Evidence lower bound (ELBO)}}+
\underbrace{KL\big(q_{\phi}(Z|X,Y)\Vert P(Z|X,Y)\big)}_{\text{Residual term}}
\label{ELBO}
\end{equation}

\noindent where $E_{Z\sim q_{\phi}}[\cdot]$ denotes expectation over $q_{\phi}(Z|X,Y)$, $KL(\cdot\Vert \cdot)$ is the non-negative Kullback-Leibler divergence of two distributions. Equation \ref{ELBO} stands at the core of variational inference theory. It decomposes the conditional likelihood into the reconstruction, regularization, and residual term, with the former two terms together coined as the ELBO.
To train our model,
we maximize the ELBO to implicitly maximize $\log p_{\theta}(Y|X)$ and minimize the residual term. 
The reason is two-fold.
First, the ELBO stands for a lower bound estimate of $\log p_{\theta}(Y|X)$, with its tightness quantified by the residual term. By maximizing the ELBO, we will approximately maximize $\log p_{\theta}(Y|X)$ while strengthening the tightness of this lower bound estimate.
The second reason is for computational convenience: by
selecting proper distribution form and adopting the \textit{reparameterization trick} \cite{kingma2013auto, rezende2014stochastic}, ELBO could be differentiable with respect to its parameters $\theta=\{\phi, \psi\}$, which allows scaling up the model to massive data via stochastic gradient optimization \cite{robbins1951stochastic, Bottou2010}.

We illustrate the latter aspect by specifying  the distribution form and the \textit{reparameterization trick}. Following \cite{kingma2013auto, rezende2014stochastic, Kihyuk2015}, we assume $q_{\phi}(Z|X,Y)$ to follow a multivariate normal distribution, with its mean vector and diagonal covariance matrix parameterized by two neural network functions,  ${\mu}_{\text{NN}_{\phi}}(X, Y)$ and ${\Sigma}_{\text{NN}_{\phi}}(X, Y)$. 
The neural network parametric forms are specified in the following section.
With this 
assumption, 
the regularization term has the following analytical form, which can be optimized using stochastic gradient optimization:
\begin{equation}
\begin{split}
KL\big(q_{\phi}(Z|X,Y)
\Vert
p(Z|X)\big) &= KL\Big(
\mathcal{N}(\mu_{\text{NN}_{\phi}},\Sigma_{\text{NN}_{\phi}})   
\Vert
\mathcal{N}(0,\mathbb{1})
\Big)
\\
&=\frac{1}{2}\Big(\text{tr}(\Sigma_{\text{NN}_{\phi}})+\mu_{\text{NN}_{\phi}}^{T}\mu_{\text{NN}_{\phi}}-k-\ln{|\Sigma_{\text{NN}_{\phi}}|}
\Big)
\end{split}
\end{equation}
Here $\mathbb{1}$ is identity matrix, $\text{tr}(\cdot)$ and $|\cdot|$ are the trace and determinant operator, $k$ is the dimensionality of the multivariate normal distributions.

To carry out gradient-based optimization for the reconstruction term, 
the \textit{reparameterization trick} is applied to substitute the latent random variable $Z\sim 
\mathcal{N}(\mu_{\text{NN}_{\phi}},\Sigma_{\text{NN}_{\phi}})$ with a $k$-dimensional standard normal distributed variable $\epsilon\sim\mathcal{N}(0,\mathbb{1})$:
$
Z=\mu_{\text{NN}_{\phi}}+\Sigma_{\text{NN}_{\phi}}\epsilon
$. 
Using this change-of-variable technique, we obtain a Monte Carlo estimate of the reconstruction term:
\begin{equation}
\begin{split}
E_{Z\sim q_{\phi}} [\log p_{\psi}(Y|X,Z)]
&=E_{\epsilon\sim
	\mathcal{N}(\pmb{0},\mathbb{1})} [\log p_{\psi}(Y|X,\mu_{\text{NN}_{\phi}}+{\Sigma}_{\text{NN}_{\phi}}\epsilon)]\\
&\triangleq\frac{1}{l}\sum_{i=1}^{l}\Big(\log p_{\psi}(Y_i|X_i,\mu_{\text{NN}_{\phi}}+{\Sigma}_{\text{NN}_{\phi}}\epsilon_{i})\Big)
\end{split}
\label{recons}
\end{equation}
Here $\triangleq$ denotes unbiased estimation, $l$ is mini-batch size.  Without loss of generality, we assume multivariate normal distribution of
$p_{\psi}(Y|X,Z)=\mathcal{N}{\Big(}\mu_{\text{NN}_\psi}(X,Z), {\Sigma_{\psi}}{\Big)}$, where $\mu_{\text{NN}_\psi}$ is a neural network function that takes into input of $X$ and $Z$ to regress $Y$, $\Sigma_{\psi}$ is the covariance matrix. With this, we have:
\begin{equation}
\resizebox{.93\linewidth}{!}{
$\underset{\phi,\psi}{\operatorname*{argmax\,}} E_{Z\sim q_{\phi}} [\log p_{\psi}(Y|X,Z)]\triangleq
\underset{\phi,\psi}{\operatorname*{argmin}}
\frac{1}{l} 
\underset{i=1}{\overset{l}{\displaystyle \sum}}
\Bigg(\Big(\mu_{\text{NN}_\psi}(X_i,\mu_{\text{NN}_{\phi}}+{\Sigma}_{\text{NN}_{\phi}}\epsilon_{i})-Y_i\Big)^{T}{\Sigma_{\psi}}^{-1}\Big(\mu_{\text{NN}_\psi}(X_i,\mu_{\text{NN}_{\phi}}+{\Sigma}_{\text{NN}_{\phi}}\epsilon_{i})-Y_i\Big)\Bigg)$
}
\label{rmse}
\end{equation}
The right-hand side of Equation \ref{rmse} is differentiable with respect to  $\{\phi, \psi\}$, allowing stochastic gradient minimization. It is worth noticing that $\Sigma_{\psi}$ balances the reconstruction term and the regularization term in composing the ELBO. Its impact is reflected in a hyperparameter $\beta$ (see Table \ref{tab:hyper} in the latter context, \cite{higgins2017beta}) in the final model.

To wrap up,
Equations \ref{ELBO}-\ref{rmse} offer an efficient variational inference framework by applying stochastic gradient ascent toward maximizing a lower bound of the conditional log-likelihood. The final loss function $\mathcal{L}$ of our learning model takes the following form:
\begin{equation}
\resizebox{.93\linewidth}{!}{
$\mathcal{L}=\frac{1}{l} 
\underset{i=1}{\overset{l}{\displaystyle\sum}}
\Bigg(\Big(\mu_{\text{NN}_\psi}(X_i,\mu_{\text{NN}_{\phi}}+{\Sigma}_{\text{NN}_{\phi}}\epsilon_{i})-Y_i\Big)^{T}{\Sigma_{\psi}}^{-1}\Big(\mu_{\text{NN}_\psi}(X_i,\mu_{\text{NN}_{\phi}}+{\Sigma}_{\text{NN}_{\phi}}\epsilon_{i})-Y_i\Big)\Bigg)
+
\frac{1}{2}\Big(\text{tr}(\Sigma_{\text{NN}_{\phi}})+\mu_{\text{NN}_{\phi}}^{T}\mu_{\text{NN}_{\phi}}-k-\ln{|\Sigma_{\text{NN}_{\phi}}|}
\Big)
$}
\label{floss}
\end{equation}
The first term encourages accurate reconstruction of the predictand using (1) the predictor information and (2) the latent variable that optimally explains the predictand. The second term forces the posterior
of the latent to fully exploit the prior distribution space, encouraging each sample from the prior to be meaningful for generating plausible predictands. This gives the premise that, without knowing $Y$ in prior, we can have good estimate of $P(Y|X)$ by sequentially sampling the conditional prior of $p(Z|X)$ and the likelihood of $p_{\psi}(Y|X,Z)$. The sampling results are thereafter used for probabilistic forecast.

\subsection{Neural network parametric form}
We specify the neural network parametric form of $\mu_{\text{NN}_\phi}(X,Y)$, $\Sigma_{\text{NN}_\phi}(X,Y)$, and $\mu_{\text{NN}_\psi}(X,Z)$ here. Neural networks are powerful function approximators that apply multiple processing layers to learn hierarchical feature representations of data \cite{lecun2015deep}. The computation units in a neural network can be flexibly organized to accommodate specific data structures and characteristics. Here we adopt convolutional neural networks (CNNs, \cite{lecun1995convolutional}) as building blocks for our model.
CNNs apply  a hierarchical set of parameterized kernels to cross-correlate with the input feature maps to explicitly exploit the \textit{translational equivariance} and \textit{compositionality} properties of the data. 
Each cross-correlation is realized using a convolution operator:
\begin{equation}
    F^{L+1}(x,y)=\sum_{i=1}^{m}\sum_{j=1}^{n}w^{L}_{i,j}F^{L}(x+i,y+j)+b^{L}
\label{conv}
\end{equation}
Here $F^{L}(x,y)$ and $F^{L+1}(x,y)$ are the $L$-th and $(L+1)$-th network layer feature vector at spatial coordinate $(x,y)$, $w^L$ and $b^L$ are the weight and bias parameter of the $L$-th layer convolution kernel, whose \textit{receptive field} size is $m\times n$.
Equation \ref{conv} enforces \textit{translational equivariance}, which tells  that,
by moving every point in the input feature map by a given distance along a given direction (translation), 
the output feature map should in response move by a same distance along a same direction. 
This property is desirable as we prefer to maintain and take use of the
spatial structure of the data
through hierarchical feature learning, and efficiently recognize features
shared across the space. 
\textit{Compositionality} refers to the property that, complicated, large scale patterns are built from simpler, small scale ones. 
In CNNs, we learn compositional features by composing several convolutional and optionally down-sampling layers, obtaining a \textit{deep neural network} that extracts local features in lower layers, while synthesizing local features for large-scale features in higher layers.
The two peculiar characteristics of CNN can find analogues in numerical geofluid dynamics solvers \cite{ruthotto2019deep}, where we approximate the local non-linear behavior of fluid dynamics in a convolution manner, and apply numerical integration to obtain compositional representations of the fluid's spatiotemporal patterns.
Recent years have witnessed increasing popularity of CNN applications in geophysical modeling \cite{pan2019improving, miao2019improving, pan2019advancing, weyn2019can, ham2019deep}.



\begin{figure}[hbt!]
\centering
\includegraphics[width=1\linewidth]{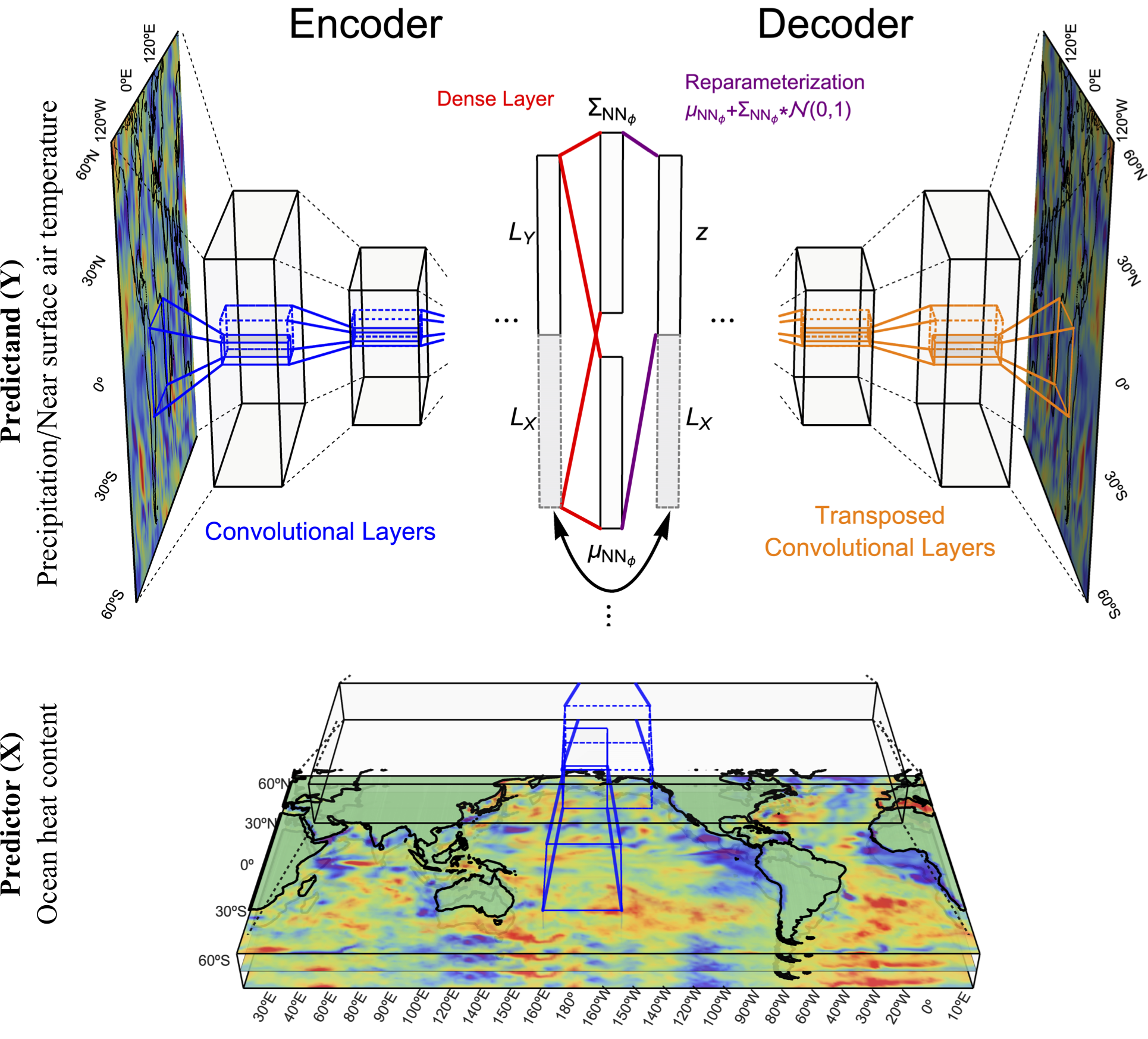}
\caption{Illustration of the Conditional Variational Auto-Encoder (CVAE) model.
The encoder, $q_{\phi}(Z|X,Y)$,  takes into input of predictor $X$ (ocean heat content) and predictand $Y$ (precipitation /near surface air temperature) to infer the conditional posterior distribution of latent variable $Z\sim 
\mathcal{N}(\mu_{\text{NN}_{\phi}},\Sigma_{\text{NN}_{\phi}})$, while the decoder, $p_{\psi}(Y|X,Z)$, takes into input of the latent variable $Z$ and predictor $X$ to generate samples of $Y$. 
$L_{X}$ and $L_{Y}$ denote the learned feature representations of the predictor $X$ and predictand $Y$ using convolutional neural networks. $L_{X}$ is shared by $q_{\phi}(Z|X,Y)$ and $p_{\psi}(Y|X,Z)$, as denoted by the black arrows. 
The blue/orange/red lines denote convolution/transposed convolution/dense layer operators in neural network modeling. The purple lines denote the reparameterization trick, showing how to generate posterior samples from $q_{\phi}(Z|X,Y)$ based on $k$-dimensional multivariate normal distributed samples and the inferred mean $\mu_{\text{NN}_{\phi}}$ and covariance matrix $\Sigma_{\text{NN}_{\phi}}$. Note that the encoder is only applied during model training.}
\label{fig1: model}
\end{figure}

Specific to our model, we apply CNNs to learn feature representations of $X$ and $Y$ in   $\mu_{\text{NN}_\phi}(X,Y)$ and $\Sigma_{\text{NN}_\phi}(X,Y)$. Meanwhile, to reduce model complexity, we re-use the CNN extracted feature of $X$ in $\mu_{\text{NN}_\psi}(X,Z)$, where
the CNN extracted feature of $X$ and the latent $Z$ are concatenated and fed into a transposed convolution network to generate high-resolution predictand from low-resolution representations. The transpose convolution is implemented by first padding an input grid's neighbours with 0, followed by a regular convolution.
This model architecture is illustrated in Figure \ref{fig1: model}.

It is worth noticing that, 
as we apply conventional CNNs on plane projections of spherical climate signals, we can not strictly guarantee \textit{translational equivariance}, due to (1) distortion at polar regions, and (2) longitudinal discontinuity at the edge of the projection.
Regarding the distortion problem, we do not consider the predictor/predictand information from polar region for the current work. This is acceptable, since most of the sources of seasonal predictability are believed to reside in the tropical oceans \cite{hall2001extratropical}.
We alleviate the impact of discontinuity by rearranging the projection of the predictor to start from 20$^{\circ}$E, which is mostly masked by non-informative 0 values from the land areas, see predictor map in Figure \ref{fig1: model}.

\subsection{Model training and forecast}

We apply 
stochastic gradient descent \cite{robbins1951stochastic} for model training. 
In preliminary experiments, we found that a joint training of $\mu_{\text{NN}_\phi}$, $\Sigma_{\text{NN}_\phi}$, and $\mu_{\text{NN}_\psi}$ toward minimizing $\mathcal{L}$ occasionally yield poor prediction performance for the validation set. We attribute this to the fact that the model may fail to fully exploit the predictor information from $X$, but rely too much on the latent information $Z$ to estimate $Y$. To have $Z$ account only for the uncertainty information of $Y$ conditioned on $X$, we develop a unique three-step training strategy.
First, we set $Z=0$, and carry out a preliminary training of $\mu_{\text{NN}_\psi}$. This is equivalent to regressing $Y$ with $X$. Second, we fix the network parameters in feature extraction of $X$, and update the rest of parameters of the model toward minimizing $\mathcal{L}$. Finally, we fine-tune all the model parameters. 




Note that the function of the encoder in our model is to find optimal transformation that maps the conditional prior of $p(Z|X)$ to the unknown distribution of $P(Y|X)$. This is achieved by (1) seeking corresponding sample space in the conditional prior of $p(Z|X)$ that optimally explains the considered sample, and (2) fully exploiting the conditional prior to have each of the sample being meaningful in generating realistic $Y$ samples.
To make probabilistic predictions without knowing $Y$ in prior, we first sample latent variable $Z$ from conditional prior of $p(Z|X)$, thereafter generate prediction $Y$  base on $p_{\psi}(Y|X,Z)$. A common strategy is to generate multiple independent samples to obtain a probability distribution as prediction. Uncertainty is inferred from the forecast spread.

\section{Experimental design and data}
\label{Section3}

As a proof-of-concept, we consider a case study of seasonal forecasting for boreal winter (October-March) mean precipitation and near-surface (2m) air temperature. Following \cite{ham2019deep} and \cite{switanek2020present}, we use the  upper ocean potential temperature profile from the previous July as predictor. 
Both predictor and predictand are regridded to $2^{\circ}\times2^{\circ}$, with a quasi-global coverage of $60^{\circ}\text{S}-60^{\circ}\text{N}$.
Other predictors and predictands can be easily incorporated into the proposed framework. 

We train our model using 
climate simulation data from GCM ensembles.
We drive the trained model using ocean reanalysis for practical forecasts for the period 1982 to 2018. The forecasts are verified against observational precipitation or temperature records. We compare the performance of our model with calibrated dynamical forecasts 
using
commonly-applied deterministic and probabilistic skill metrics. The applied datasets, dynamical forecast baselines,
data processing methods, model architecture, and skill evaluation approaches are described below. 

\subsection{Data}
\subsubsection{Climate simulation} 
The climate simulation data for model training are obtained from the Phase 5 \cite{taylor2012overview} and Phase 6 \cite{eyring2016overview} of the Coupled Model Intercomparison Project (CMIP5/CMIP6). To achieve sample diversity and support the training of large-scale statistical models, climate simulations from various forcing scenarios are considered, including historical climate change simulations (in response to natural and human forcings), pre-industrial control simulations (representing natural climate variability that occurs in absence of external forcing agents), 1 percent/yr CO2 increase simulations, abrupt 4xCO2 simulations, and projections of future climate change guided by various representative greenhouse gas concentration pathways (for the CMIP5 projections) or shared socioeconomical pathways (for the CMIP6 projections). 
To date, we have collected 127,137 years of samples, including 62,155 samples from CMIP5 (Appendix Table 1), and 64,982 samples from CMIP6 (Appendix Table 2), generated by GCMs developed by 12 research and forecast centers. 
Each sample consists of a predictor-predictand pair, where the predictors are July ocean potential temperature profile sampled at 15 ocean top layers that hold seasonal variability (5m, 15m, 25m, 35m, 45m, 55m, 75m, 100m, 125m, 150m, 175m, 200m, 225m, 275m, and 300m), and the predictands are the following October-March mean precipitation and near-surface (2m) air temperature.

\subsubsection{Observational references}
We use the  ECMWF ocean reanalysis system 5 (ORAS5, \cite{zuo2017new}) July ocean potential temperature data as input to force our trained model for practical forecasts. To verify models' prediction skills, we compare models' forecasts with  precipitation observational records from the
Global Precipitation Climatology Project (GPCP)  \cite{huffman1997global}, and 2m air temperature reanalysis records from the fifth generation of the  European Centre for Medium-Range Weather Forecasts (ECMWF) atmospheric reanalyses (ERA5, \cite{hersbach2020era5}). 
The boreal winter mean observation are obtained by averaging the monthly data from October to March for each year from 1982 to 2018. 


 


\subsubsection{Dynamical forecasts}
We compare our model's forecast performance with the North American Multi-model Ensemble (NMME) Phase-II retrospective forecast \cite{kirtman2014north}. We consider four NMME dynamical seasonal forecast systems, namely the Community Climate System Model Version 4  (CCSM4, \cite{jahn2012late}) developed by National Center for Atmospheric  Research, 
the third  and fourth generation of Canadian Coupled Global Climate Model (CanCM3/CanCM4, \cite{merryfield2013canadian}) developed by the Canadian Centre for Climate Modelling and Analysis, and the Forecast-oriented Low Ocean Resolution Model (FLOR-B01, \cite{msadek2014importance}) developed by
the Geophysical Fluid Dynamics Laboratory.  Each model has 10 or more ensemble members. For consistency, we select only 10 ensemble members from each model.
All forecasts have been bias corrected using cross-validation \cite{kirtman2009multimodel}.
We consider the forecasts that are initialized in July from 1982 to 2018, and use their regridded October to March mean forecast as our baseline. 

\subsection{Data processing}
Both predictors and predictands are spatial maps, regridded to a common $2^{\circ}\times2^{\circ}$ resolution using the nearest neighbor method. The predictor variable of ocean potential temperature at all depth layers is normalized using a common min-max normalization by subtracting the min and dividing by (max-min). The min and max scaling factors are set to 272.15 K and 310.15 K. Data for the land areas are masked by 0. The predictor is projected to start from 20$^{\circ}$E, 
which is mostly masked by non-informative 0 values from the land areas.
The predictand variables are normalized at each grid point by subtracting the spatial mean, and dividing by the standard deviation of the observational records from 1982 to 2018 at grid scale. 

We divide the climate simulation data into the training, validation, and test sets. The test sets are composed of the CMIP5 and CMIP6 historical climate simulations performed with each GCM from 1982 to 2007/2013. These data are selected to maximally match the practical forecast verification period. 90\% and 10\% of the rest data are shuffled and allocated to the training and validation sets.


\subsection{Model architecture, hyperparameters and training}
The structure, hyperparameters, and training details of the applied model is illustrated in Table \ref{tab:hyper}.
We use the Adaptive Moment Estimation (Adam, \cite{kingma2014adam}) stochastic gradient descent for training. 
For each of the three steps in training,
we set the maximum number of training epochs to 200, and employ an early stopping strategy: training is stopped when the relative decrease of the validation loss is less than 0.1\% for 5 consecutive training epochs.


\begin{table}
    \centering
    \caption{Structure$^{\dagger}$, hyperparameters, and training details of the model}\resizebox{!}{0.3\textwidth}{
\begin{threeparttable}
    \begin{tabular}{cccc}
    \hline
    \multicolumn{4}{c}{\includegraphics[width=.8\textwidth]{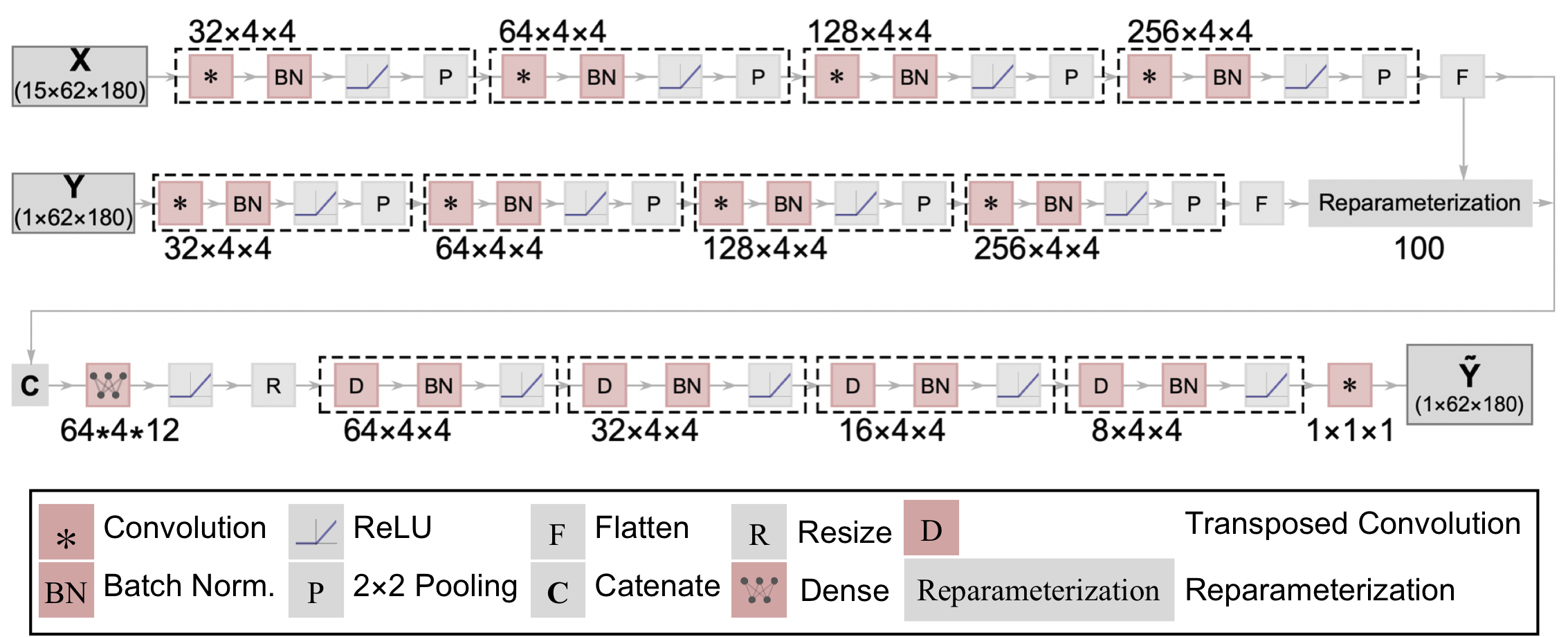}}\\
        \hline
    \multicolumn{3}{c}{Hyperparameter}
    &Considered options\\
    \hline
     \multirow{7}{*}{\rotatebox[origin=c]{0}{Encoder}}&\multirow{3}{*}{\rotatebox[origin=c]{0}{${L_X}^{\dagger\dagger}$}}&Receptive field ($m\times n$)&$(3\times3)$, $\underline{(4\times4)}^{\ddagger}$
     \\
     &&Convolution channel&${\underline{16\times2^{\mathcal{D}_{E}}}}^{\ddagger\ddagger}$
     \\
     &&Number of convolution blocks$^{\mathsection}$ & 3,  \underline{4}
     \\
     
     \cline{3-4}
      
     &\multirow{3}{*}{\rotatebox[origin=c]{0}{$L_Y^{\dagger\dagger}$}}&Receptive field ($m\times n$)&
     $(3\times3)$, $\underline{(4\times4)}$
     \\
     
     &&Convolution channel &$\underline{16\times2^{\mathcal{D}_{E}}}$$^{\ddagger\ddagger}$
     \\
     
     &&Number of convolution blocks & 3,  \underline{4}\\
     \cline{3-4}
     &$z$&Dimension of $z$ ($k$)
     &50, \underline{100}, 200
      \\
     \cline{2-4}
      
     \multirow{4}{*}{\rotatebox[origin=c]{0}{Decoder}}&\multicolumn{2}{c}{Initial dimension}&
     $32\times4\times12$, $\underline{64\times4\times12}$
     \\
     &\multicolumn{2}{c}{Receptive field ($m\times n$)}& $\underline{(4\times4)}$
    
     \\
     
     &\multicolumn{2}{c}{Transposed convolution channel}&$\underline{128\times2^{-\mathcal{D}_{D}}}$$^{\ddagger\ddagger}$
     \\
     
     &\multicolumn{2}{c}{Number of transposed 
     convolution blocks} & \underline{4}
     \\
      \cline{1-4}
     \multirow{5}{*}{\rotatebox[origin=c]{0}{Training}}&\multicolumn{2}{c}{Optimizer}&Adam \small{($\beta_1=0.9, \beta_2=0.999, \epsilon=10^{-5}$)}\\
     
     &\multicolumn{2}{c}{Learning rate}&$10^{-3}$, \underline{$10^{-4}$}, $10^{-5}$\\
      &\multicolumn{2}{c}{Batch size}& \underline{64}\\
      &\multicolumn{2}{c}{$\beta^{\mathsection\mathsection}$}&1,  $\underline{5}$, 10\\
    \hline
    \end{tabular}
    \begin{tablenotes}
    \item[$\dagger$] Structure of the model is shown in the first row. The model has 9,211,593 trainable parameters.\\ 
    \item[$\dagger\dagger$] $\mathcal{L}_{X}$ and $\mathcal{L}_{Y}$ are the convolutional neural network feature representations of predictor $X$ and  predictand $Y$.\\
    \item[$\ddagger$] The selected hyperparameters are labeled with underline.\\
    \item[$\ddagger\ddagger$] $\mathcal{D}_E$ and $\mathcal{D}_D$ denote the depth of the convolution blocks. We sequentially double/halve the convolution channel in the encoder/decoder.\\
    \item[$\mathsection$]Each convolution block is composed of a convolutional layer, rectified linear activation ($\text{ReLU}(x)=max(0,x)$), batch-normalization\cite{ioffe2015batch}, and $2\times2$ maximum pooling.\\
     \item[${\mathsection\mathsection}$]$\beta$ balances the reconstruction and regularization terms in the evidence lower bound equation.\\
    \end{tablenotes}
\end{threeparttable}
}
\label{tab:hyper}
\end{table}

\subsection{Skill metrics}

We use the Pearson correlation coefficient score ($r$) and normalized root mean square error score (NRMSE) to measure the deterministic forecast skill of the ensemble mean predictions.
In addition, we use the area under the
ROC (Receiver Operating Characteristics) 
curve score (AUC, \cite{kharin2003roc})
and the continuous ranked probability skill score (CRPSS, \cite{hersbach2000decomposition}) to assess the probabilistic skill of the full ensemble of forecasts. 
A summary of the skill metrics, their computation methods, and ranges is given in Table \ref{tab:skillmetrics}.

Regarding the significance tests for the skill differences between models, we notice that
the seasonal predictability exhibits strong annual variability, inhibiting the establishment of asymptotic distribution forms of models' skill score statistics. Thereafter, we apply bootstrap
for significance test of models' forecast skill differences.
We first generate 100
bootstrap samples of forecasting cases using random sampling with replacement. Thereafter, we calculate the ratio of forecast skill scores between two models for each bootstrap sample.
The confidence interval is thereafter determined using a percentile method. 
All tests are single-side at confidence level of 95\%.

\begin{table}
    \centering
    \caption{Skill metrics for forecast evaluation}
    
\begin{threeparttable}
    \begin{tabular}{ccccccc}
    \hline
    Category&Skill metrics&Calculation&Range$^{\dagger}$\\
    \hline
     \multirow{4}{*}{Deterministic} &Pearson correlation coefficient &
     \multirow{2}{*}{$r^{\dagger\dagger} = \frac{{}\sum_{i=1}^{n} (y_\text{simu}^i - \overline{y_\text{simu}})(y_\text{obser}^i - \overline{y_\text{obser}})}
{\sqrt{\sum_{i=1}^{n} (y_\text{simu}^i - \overline{y_\text{simu}})^2}\sqrt{(y_\text{obser}^i - \overline{y_\text{obser}})^2}}$}&
  \multirow{2}{*}{(0,1)}\\
    &($r$)\\
     &Normalized root mean square error &
     \multirow{2}{*}{$\text{NRMSE}=1-\frac{\sqrt{\sum_{i=1}^{n} (y_\text{obser}^i-y_\text{simu}^i)^2}}{\sqrt{\sum_{i=1}^{n} (y_\text{obser}^i - \overline{y_\text{obser}})^2}}$}
     &\multirow{2}{*}{(0,1)}
     \\
     &(NRMSE)\\
     \hline
     \multirow{6}{*}{Probabilistic}&&Step 1: Construct the
     ROC$^{\ddagger}$ curve by&\multirow{4}{*}{(0.5,1)}\\
     &Area under the curve& drawing hit rate against false alarm rate  \\
     &(AUC)& for predictions exceeding the ten deciles.\\
     &&Step 2: Compute area under the curve.\\
     
      & Continuous ranked probability skill score &
      \multirow{2}{*}{$\text{CRPSS}^{\ddagger\ddagger}=1-\frac{\sum_{i=1}^{n}\int_{-\infty}^{\infty}[P_{\text{simu}}(y_i)-P_{\text{obser}}(y_i)]^2dy_i}{\sum_{i=1}^{n}\int_{-\infty}^{\infty}[P_{\text{ref}}(y_i)-P_{\text{obser}}(y_i)]^2dy_i}$}&\multirow{2}{*}{(0,1)}\\
      &(CRPSS)&&\\
    \hline
    \end{tabular}
    \begin{tablenotes}
    \item[$\dagger$] Min/max value indicates no skill/perfect skill. 
    \newline
    \item[$\dagger\dagger$]$n$ is sample size, $y_\text{simu}^i$, $y_\text{obser}^i$
    are individual simulation/observation samples indexed by $i$. Overbar denotes climatology mean.
    \newline
    \newline
    \item[$\ddagger$] Receiver Operating Characteristics. See appendix for detailed explanation. 
    \newline
    \item[$\ddagger\ddagger$] $P_{\text{simu}}$ /$P_{\text{obser}}$ is the cumulative probability distribution function of the simulation/observation. 
    $P_{\text{obser}}(y)= 
    \begin{cases}
    0,& \text{if } y<y^{\text{obser}} \\
    1,              & \text{otherwise}
    \end{cases}$. $P_{\text{ref}}$ is the cumulative probability distribution of climatology records. 
    See \cite{pan2019precipitation} for  computation details. 
    \end{tablenotes}
\end{threeparttable}
\label{tab:skillmetrics}
\end{table}

\section{Results}
\label{Section4}

\begin{figure*}
\centering
\includegraphics[width=\linewidth]{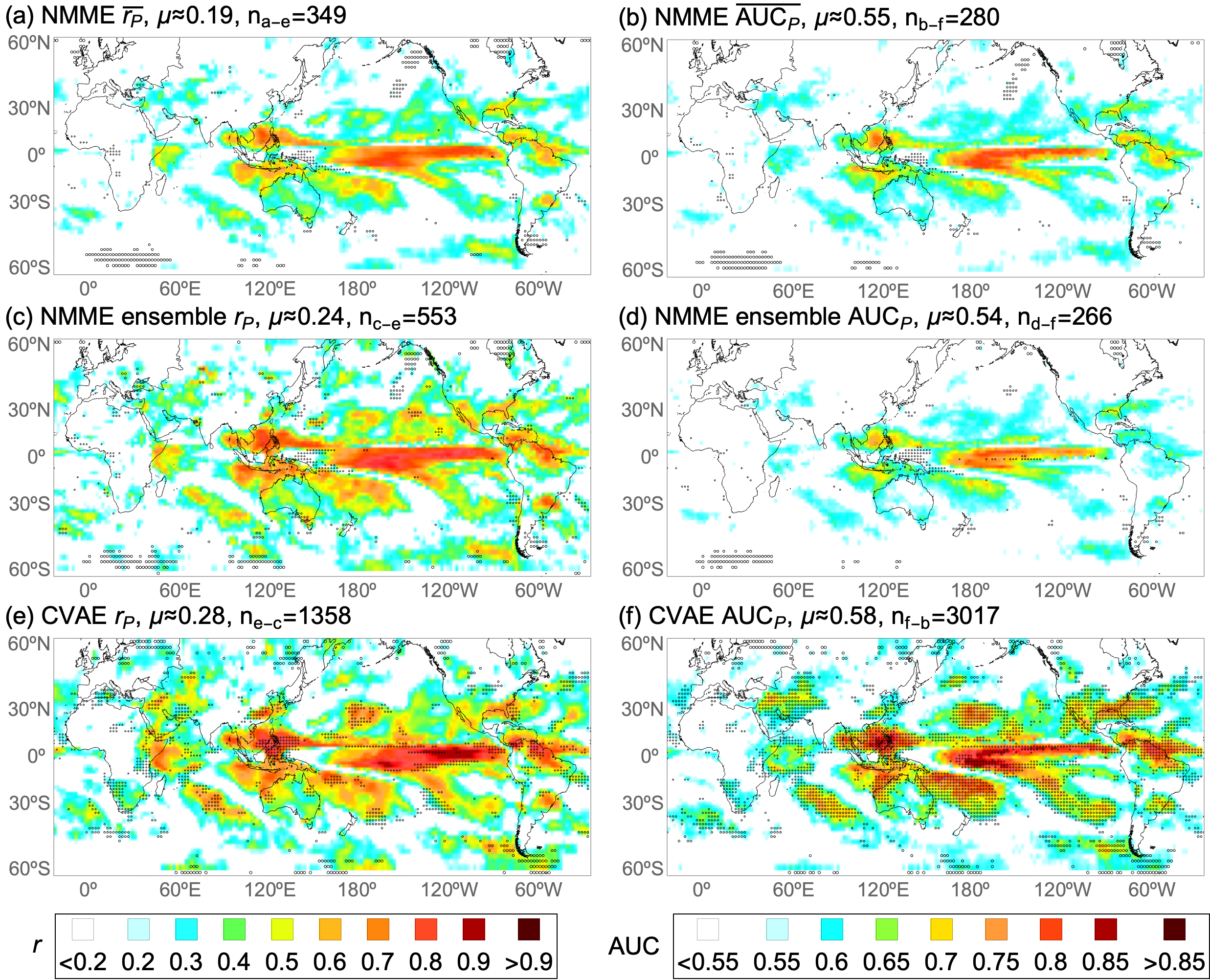}
\caption{Forecast skill for October-March mean precipitation using models starting in July for 1982 to 2018 at $2^{\circ}\times2^{\circ}$ grid scale.
(a, b), the intra-model mean $r$ and AUC skill score of four NMME models.
(c, d), the $r$ and AUC skill score of the NMME four-model ensemble.
(e, f), the $r$ and AUC skill score of the CVAE model.
Stippling in (a-d) shows locations where the NMME forecasts show significant (95\% confidence level, same for the rest tests) higher skill than CVAE forecasts. 
Stippling in (e) shows locations where the CVAE forecasts show significant higher $r$ score than NMME ensemble forecasts. 
Stippling in (f) shows locations where the CVAE forecasts show significant higher AUC score compared to the mean AUC score of four NMME models. 
The spatial average skill score and the number of stippling are denoted by $\mu$ and $n$ in each sub-figure.}
\label{fig1: precipitation}
\end{figure*}

\begin{figure*}
\centering
\includegraphics[width=\linewidth]{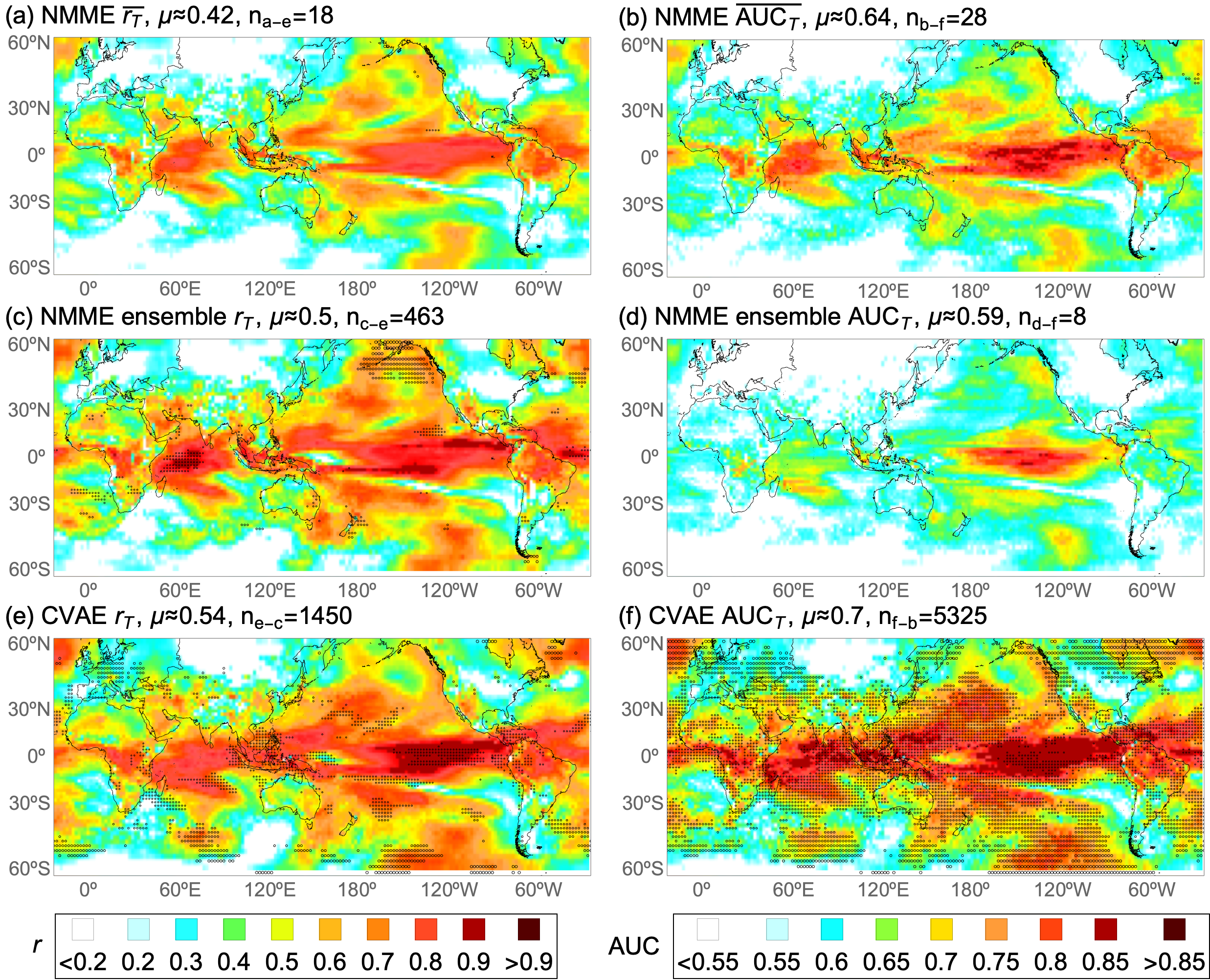}
\caption{Similar as Figure \ref{fig1: precipitation} but for 2m air temperature forecast.}
\label{fig2: temperature}
\end{figure*}

We first compare the performance of the NMME dynamical forecast systems and the CVAE model for  predicting the
observed October-March mean precipitation (Fig.~\ref{fig1: precipitation}) and 2m air temperature (Fig.~\ref{fig2: temperature}). All forecasts are made from the previous July, with forecast lead time of 3-8 months. Evaluations are carried out at $2^{\circ}\times2^{\circ}$ grid scale for the period of 1982 to 2018. For the NMME forecasts, we consider single model 10-member based forecast, the intra-model mean skill of the four NMME models are shown in Fig.~\ref{fig1: precipitation}a,b and \ref{fig2: temperature}a, b. We also evaluate the four-model ensemble ($4\times10$-member) forecast, results are shown in Fig.~\ref{fig1: precipitation}c, d and \ref{fig2: temperature}c, d. For the CVAE, we 
apply a 1000 ensemble member forecast created by sampling the latent variable $Z$. The results are shown in Fig.~\ref{fig1: precipitation}e, f and \ref{fig2: temperature}e, f. 
In these evaluations,
we show the \textit{r} skill score of the ensemble mean forecasts (left column of Fig.~\ref{fig1: precipitation} and \ref{fig2: temperature}), as well as the AUC score
that summarizes ensemble forecasts' hit rates and false alarm rates for a range of probability thresholds (right column of Fig. ~\ref{fig1: precipitation} and  \ref{fig2: temperature}). Stippling denotes grids where the NMME and CVAE forecasts show significant skill difference at 95\% confidence level. The spatial average skill score and the number of stippling gridboxes are used to roughly summarize models' forecast capability. Evaluation using NRMSE and CRPSS shows similar patterns, and is given in the Appendix.

Compared to the NMME dynamical forecasts,
the CVAE model exhibits similar skill distribution pattern,  but achieves an overall higher level of deterministic and probabilistic skills. 
Specific for precipitation forecast (Fig.~\ref{fig1: precipitation}), 
the CVAE forecasts 
show significantly higher $r$ and AUC skill score in Eurasia, Eastern and Southern Africa, tropical Pacific, Southern U.S., and Amazon (Fig.~\ref{fig1: precipitation}e, f), 
compared to either the intra-model mean skill score of the NMME forecasts (Fig.~\ref{fig1: precipitation}a, b), or the skill score of the NMME multi-model ensemble  forecast (Fig.~\ref{fig1: precipitation}c, d).
While the multi-model ensemble can considerably enhance the $r$ skill score of dynamical forecasts (Fig. ~\ref{fig1: precipitation}a, c),
reducing its gap with CVAE forecasts (Fig.~\ref{fig1: precipitation}e), 
we notice that the AUC scores of the multi-model ensemble is worse than individual models for most regions, which may due to the climatology mismatch of the considered dynamical forecast models.

For 2m air temperature forecast (Fig.~\ref{fig2: temperature}), both NMME and CVAE forecasts achieve higher skill scores compared to precipitation forecasts. 
The CVAE model demonstrates particularly strong skill in predicting the temperature variation for the eastern Pacific ENSO region, with $r>0.9$ (Fig.~\ref{fig2: temperature}e) and AUC$>0.85$ (Fig.~\ref{fig2: temperature}f). 
Also, the CVAE model shows significantly higher prediction skill for regions where the NMME skill is limited, such as western Eurasia and central North America.

It is worth noticing that the skills of the CVAE model shown here are derived purely from GCM simulations,
no observational data has been used for training the CVAE model, ensuring that we run a small risk of ``overfitting'',
in the sense that the CVAE predictions reflect quirks of the climate simulation data, rather than regularities generalizable to the real-world climate system.
Based on these results, we have shown that
the CVAE model offers a competitive benchmark for both deterministic and probabilistic seasonal forecasting.

\section{Discussion}
\label{Section5}
\subsection{Forecast diagnosis}
\label{diagnosis}


How can we outperform GCM-based dynamical seasonal forecast using GCM simulation information only? What are the implications for diagnosing and improving dynamical seasonal forecast? 
We answer these questions by contrasting
a same-GCM supported dynamical forecast and CVAE forecast, applied to either the real-world climate system
or the GCM simulation ``model world''.
Corroborated by multi-GCM supported forecasting results in Fig.~\ref{fig1: precipitation} and Fig.~\ref{fig2: temperature}, we reveal the forecasting barriers posed by GCM initialization and formulation deficiencies, as well as this GCM's predictability limit. 

\begin{figure}[ht]
\centering
\includegraphics[width=\linewidth]{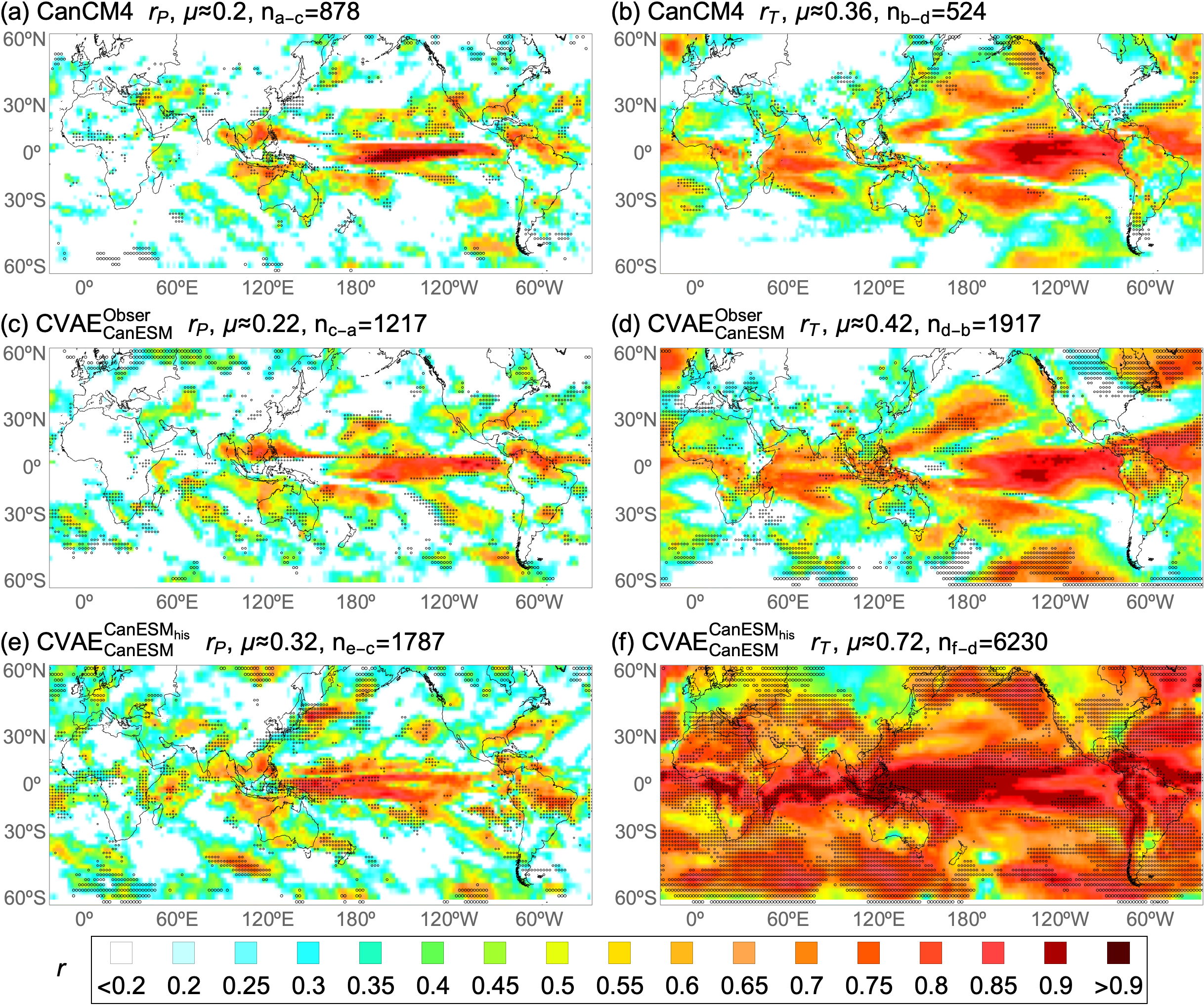}
 \caption{Correlation coefficient ($r$) skill score of CanCM4 and CanESM-supported CVAE forecasts.
 (a, b), $r$ skill score for CanCM4 precipitation/temperature forecast.
 (c, d), $r$ skill score for CanESM-supported CVAE precipitation/temperature forecast.
 (e, f), $r$ skill score for CanESM-supported CVAE forecast applied to CanESM historical simulations. 
 Stippling in (a, b) denotes locations where the CanCM4 forecasts show significant (95\% confidence level, same for the rest tests) higher skill than corresponding CVAE forecasts. 
 Stippling in (c, d) denotes locations where the CVAE forecasts show significant higher skill than CanCM4 forecasts. 
 Stippling in (e, f) denotes locations where the CVAE forecasts applied to the CanESM historical simulation show significant higher skill than the CVAE forecasts applied to real-world climate system.
 The spatial average skill score and the number of stippling are labeled.
}
\label{fig4: finetune}
\end{figure}

We choose the Canadian Earth System Model (CanESM) and its corresponding seasonal forecast system CanCM4 as an example. 
We consider the following two models: (1) the CanCM4 dynamical forecast system, and (2) a new CVAE model trained with the 12,253 years of the CanESM simulation data ($\text{CVAE}_\text{CanESM}$).
We apply $\text{CVAE}_\text{CanESM}$ 
for forecast
in (1) the observed real-world climate system, and (2) the ``model world'' realized by the $\text{CanESMv5}$  historical simulation. These forecasts are denoted as $\text{CVAE}_{\text{CanESM}}^\text{Obser}$ and $\text{CVAE}_{\text{CanESM}}^{\text{CanESM}_{\text{his}}}$, respectively.
Results measured using $r$ skill score are shown in Figure \ref{fig4: finetune}. Comparing the CanCM4 and $\text{CVAE}_{\text{CanESM}}^{\text{Obser}}$ highlights potential initialization errors, while contrasting $\text{CVAE}_{\text{CanESM}}^\text{Obser}$, $\text{CVAE}_{\text{CanESM}}^{\text{CanESM}_{\text{his}}}$ and $\text{CVAE}_{\text{Ensemble}}^\text{Obser}$ (our CVAE model based on multiple CMIP5 and CMIP6 simulations)   highlights the model formulation deficiency and predictability limits. 





\subsubsection{Initialization error}
Since both the CanCM4 and the $\text{CVAE}_{\text{CanESM}}$  are based on the same  
GCM, their performance difference in practical forecast should
largely depend upon how they make use of this GCM. 
In CanCM4 forecasts, we run simulations starting from an ensemble of initial states that agree with observational constraints. 
In $\text{CVAE}_{\text{CanESM}}^\text{Obser}$  forecasts, we make use of the seasonal variability signal learned from continuous runs of CanESM simulations. 
Although the probabilistic relationship learned by the $\text{CVAE}_{\text{CanESM}}$  may be oversimplified or misrepresented, the benefit of this simplification lies in that, it allows bypassing iterative state updating in forecast, thereafter greatly alleviate the drift problem in initializing dynamical forecasts.

The discussion above tells that,
the skill difference between the CanCM4 and the $\text{CVAE}_{\text{CanESM}}^\text{Obser}$  reflects
the following tug-of-war: Does the initialization error in dynamical forecasts surpass the error of the CVAE model resulting from oversimplifying or misrepresenting the dominant factor of seasonal variability?
For precipitation forecasts (Fig.~\ref{fig4: finetune}a, c),  while
CanCM4 demonstrates significant better $r$ skill score for certain regions, such as tropical and western mid-latitude Pacific, there are many other regions where $\text{CVAE}_{\text{CanESM}}^\text{Obser}$  obtains significant better $r$ skill score, such as northern Eurasia, central U.S., and Southern Hemisphere mid latitude.
For 2m air temperature forecasts (Fig. ~\ref{fig4: finetune}b, d), compared to CanCM4, $\text{CVAE}_{\text{CanESM}}^\text{Obser}$  achieves significant better $r$ skill score for a broader range of regions, including both continental and ocean parts. 
These results indicate that CanCM4 has not fully exploited the seasonal prediction capacity of its GCM. 
We believe this skill lagging of the CanCM4 compared to $\text{CVAE}_{\text{CanESM}}^\text{Obser}$  offers a lower-bound estimate of unrealized prediction capacity of the GCM due to initialization deficiencies. 

\subsubsection{Model formulation deficiency and predictability limit}
It has become a conventional practice to reveal the defect of a single GCM by comparing its forecast with forecast from multi-GCM ensembles.
In dynamical forecast, this comparison  reflects a compound impact of initialization and GCM formulation deficiencies. 
The previous section has demonstrated that the initialization deficiencies can be largely alleviated using the CVAE methodology.
Regarding the GCM formulation deficiencies, it is beneficial to further clarify whether they originate from a biased representation of the seasonal variability signal, or  an underestimation of this signal. We 
shed light on this problem 
by contrasting
the following three CVAE forecasts: (1)
$\text{CVAE}_{\text{Ensemble}}^\text{Obser}$, 
the multi-GCM ensemble based CVAE forecast applied to the observed real world (Fig.~\ref{fig1: precipitation}e and 
\ref{fig2: temperature}e), 
(2) $\text{CVAE}_{\text{CanESM}}^\text{Obser}$, 
the single GCM (CanESM) based CVAE forecast applied to the observed real world (Fig.~\ref{fig4: finetune}c, d), 
and (3) $\text{CVAE}_{\text{CanESM}}^{\text{CanESM}_{\text{his}}}$, the single GCM  based CVAE forecast applied to this GCM's historical simulation ``model world'' (Fig.~\ref{fig4: finetune}e, f). 
The results  reveal the achieved and potentially achievable seasonal forecast skill of the considered GCM. 

\begin{figure}[hbt!]
\centering
\includegraphics[width=\linewidth]{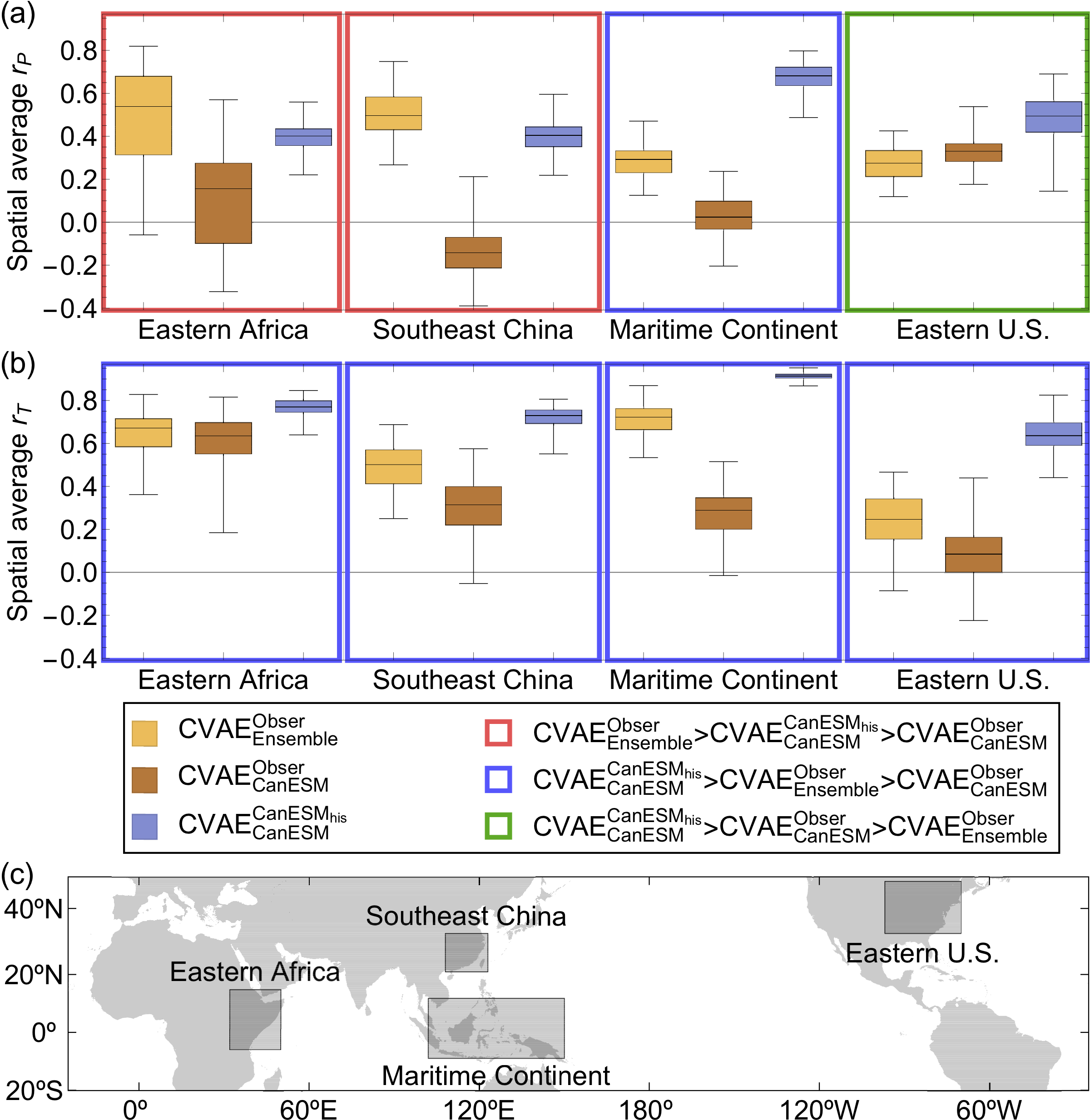}
\caption{Bootstrap distribution of spatial average $r$ skill scores for three CVAE forecasts in selected regions. The three CVAE forecasts are 
(1) $\text{CVAE}_{\text{Ensemble}}^\text{Obser}$, 
the multi-GCM ensemble based CVAE forecast applied to the observed real world,
(2) $\text{CVAE}_{\text{CanESM}}^\text{Obser}$, 
the single GCM (CanESM) based CVAE forecast applied to the observed real world, 
and (3) $\text{CVAE}_{\text{CanESM}}^{\text{CanESM}_{\text{his}}}$, the single GCM  based CVAE forecast applied to this GCM's historical simulation ``model world''.
The four selected regions are where the $\text{CVAE}_{\text{Ensemble}}^\text{Obser}$ and $\text{CVAE}_{\text{CanESM}}^\text{Obser}$ forecasts demonstrate significant skill differences.
Bootstrap samples are generated by randomly sampling the forecast cases with replacement.
Results are labeled with colored rectangles depending on the ranking of $r$ for the three CVAE forecasts.
(a), bootstrap distribution of spatial average $r$ skill scores for  precipitation forecast represented using box charts. The lines from bottom to top in a box show the min, 25\% quantile, median, 75\% quantile, and max of the samples. 
(b), results for 2m temperature.
(c), the selected regions.}
\label{fig5: region}
\end{figure}

While the $\text{CVAE}_{\text{Ensemble}}^\text{Obser}$ and $\text{CVAE}_{\text{CanESM}}^\text{Obser}$ forecasts
show considerably similar spatial distribution of skill (Fig. ~\ref{fig1: precipitation}e and Fig. ~\ref{fig4: finetune}c, Fig. ~\ref{fig2: temperature}e and Fig. ~\ref{fig4: finetune}d),
we focus on the hot spots where these two forecasts show significant skill differences. 
Figure \ref{fig5: region} shows the bootstrap distribution of spatial average $r$ skill scores for four of these hot spots, namely Eastern Africa, Southeast China, Maritime Continent, and Eastern U.S.  We distinct three configuration cases based on the ranking of the spatial average $r$ skill score achieved by these CVAE forecasts. 

\begin{enumerate}
    \item $\text{CVAE}_{\text{Ensemble}}^\text{Obser}>\text{CVAE}_{\text{CanESM}}^{\text{CanESM}_{\text{his}}}>\text{CVAE}_{\text{CanESM}}^{\text{Obser}}$. This ranking of $r$ skill score (red rectangle delineated box charts in Fig.~\ref{fig5: region}) suggests that, compared to the real-world climate system, the CanESM model alone fails to develop a strong enough constraint posed by the prediction sources in its simulations. This happens for precipitation forecast in Eastern Africa and Southeast China. For both cases, this overly weak constraint of CanESM is meanwhile biased when applied for practical forecast, given the negative $r$ skill score  of the $\text{CVAE}_{\text{CanESM}}^{\text{Obser}}$ forecast.
    \item $\text{CVAE}_{\text{CanESM}}^{\text{CanESM}_{\text{his}}}>\text{CVAE}_{\text{Ensemble}}^\text{Obser}>\text{CVAE}_{\text{CanESM}}^{\text{Obser}}$. This ranking of $r$ skill score (blue rectangle delineated box charts in Fig.~\ref{fig5: region}) represents the most common case. It suggests that, compared to the real-world climate system, the CanESM model develops a stronger constraint posed by the prediction sources in its simulations. However, this strong seasonal variability signal does not apply well to the real-world climate system, compared to the seasonal variability signal revealed by the multi-GCM ensemble. 
    \item $\text{CVAE}_{\text{CanESM}}^{\text{CanESM}_{\text{his}}}>\text{CVAE}_{\text{CanESM}}^{\text{Obser}}>\text{CVAE}_{\text{Ensemble}}^\text{Obser}$. This ranking of $r$ skill score (green rectangle delineated box charts in Fig. ~\ref{fig5: region}) tells that, compared to the multi-GCM ensemble, the considered CanESM is advantageous in representing the seasonal variability signal, and achieves a better forecast skill for the selected region.
\end{enumerate} 

The discussion above suggests that, 
for the considered CanESM example, 
there are regions where the GCM develops overly weak and biased constraints on precipitation variability (red rectangle delineated box charts in Fig.~\ref{fig5: region}a). 
We believe these are hot spots worth
close and urgent attention in model development.
Besides these regions,
the prediction sources from upper ocean thermal profile usually pose a stronger constraint on seasonal variability in the GCM simulation
than in the real-world climate system, especially for temperature (Fig.~\ref{fig4: finetune}f and Fig.~\ref{fig5: region}b). 
This strong signal reflects the potentially achievable forecast skill of the considered GCM. While this predictability signal may vary with both externally forcings and internal climate variability, we examine how this signal depends upon the dominant seasonal variability mode of El Ni\~{n}o/Southern Oscillation (ENSO) in the latter section, 
and leave the discussion for the impact of externally forcings in future works.
Overall, the comparison results here help to reveal the GCM formulation deficiencies, and pinpoint the directions toward forecast improvement.

\subsection{Identifying sources of predictability}
Identifying the sources of predictability  helps specifying the key climate processes that offer predictability. 
In dynamical simulation, this task often requires costly computation as we examine GCM's responses to changes from many potentially crucial factors. Here, we alleviate this computation burden by analyzing the CVAE model instead. The analysis in turn tells if the CVAE model applies physically-meaningful factors in its forecast.

\begin{figure}[ht]
\centering
\includegraphics[width=\linewidth]{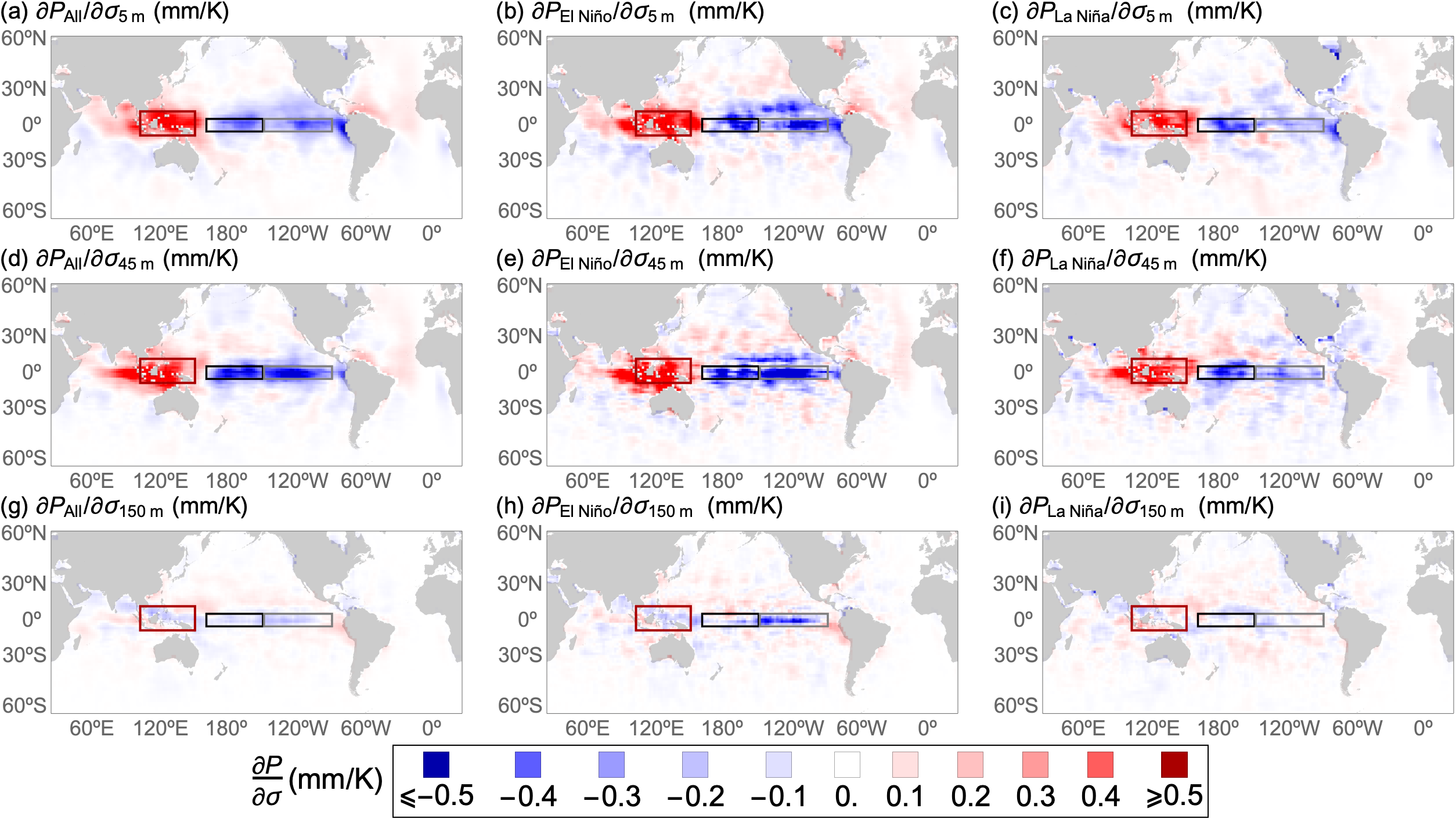}
\caption{Saliency map for precipitation prediction in the Maritime Continent. (a), change of CVAE predicted October-March precipitation in the Maritime Continent due to  one standard deviation increase of potential temperature during the previous July at 5m ocean depth. Results are calculated for each year from 1982 to 2017, the average values are plotted. 
(b), similar as (a) but for strong 
Ni\~{n}o years (Year 1987, 1997, and 2015, when the $\text{Ni\~{n}o3.4}$ index is larger than 1).
(c), similar as (a) but for strong La Ni\~{n}a years (Year 1988,1999, and 2010, when the $\text{Ni\~{n}o3.4}$ index is smaller than -1).
(d-f) and (g-i), similar as (a-c) but for 45m and 150m ocean depth.
The red, black and grey polygon denote the Maritime Continent, the Ni\~{n}o4 and Ni\~{n}o3 region.}
\label{fig6:saliency}
\end{figure}

The technique we apply here is called \textit{saliency map} analysis, which shows how each portion of the predictor influences the forecast by visualizing the neural network gradient. In other words, it is a visualization tool highlighting the geographical aeras that have influenced the forecast the most.   
Consider an example case of precipitation forecast in the Maritime Continent, 
we calculate the change of CVAE ensemble-mean precipitation forecast for this region due to a one standard deviation increase of potential temperature during the previous July at each location and ocean depth. Results for all the forecasting years (1982-2017), the strong El Ni\~{n}o years ($\text{Ni\~{n}o3.4}$ index in July is larger than 1), and the strong La Ni\~{n}a years ($\text{Ni\~{n}o3.4}$ index in July is smaller than -1) are averaged and plotted for three representative ocean depth layers (5m, 45m, and 150m) in Fig.~\ref{fig6:saliency}. 

The zonal dipole patterns in Fig.~\ref{fig6:saliency}
suggest that, the CVAE precipitation forecast in the Maritime Continent is negatively influenced by the July east-central Equatorial Pacific top-layer ocean potential temperature, and positively influenced by the July local top-layer ocean potential temperature. 
These patterns faithfully reflect the dominant role of the Walker circulation. A strong Walker circulation, arising from an enhanced ocean thermal gradient in the equatorial tropical Pacific, features strengthened convection and abundant precipitation in its rising branch around the Maritime Continent. Contrarily, a weak Walker circulation, caused by anomalous ocean warming in the east-central Pacific, shifts its rising branch to the east, and introduces anomalous sinking over the Maritime Continent, suppressing precipitation formation.

The well-pronounced impact at 45m ocean depth (Fig.~\ref{fig6:saliency}d-f)
highlights the role of below-surface ocean thermal structure for leading and sustaining the sea surface and atmosphere circulation variations at seasonal lead time. Furthermore, the comparison between saliency responses in El Ni\~{n}o years (Fig.~\ref{fig6:saliency}b, e, h) and La Ni\~{n}a years (Fig.~\ref{fig6:saliency}c, f, i) illustrates the asymmetric impact of ENSO, namely precipitation during El Ni\~{n}o is more sensitive to ocean thermal variations in the eastern Ni\~{n}o 3 region (grey rectangle box), especially for the deeper ocean, while precipitation during La Ni\~{n}a is more sensitive to ocean thermal variations in the central Ni\~{n}o 4 region (black rectangle box) and less influenced by the deeper ocean. 

These findings, corroborated by existing observation and simulation studies \cite{TokinagaEtAl2012}, confirm that the CVAE model leverages physically-meaningful patterns in its predictions. This \textit{saliency map} analysis enables efficient identification of key prediction sources. 

\subsection{Impact of ENSO on predictability}
\label{dfd}

Seasonal forecasts starting from different 
climate initial states may demonstrate distinct levels of skill, which has been partially attributed to the impact of internal climate variability on predictability \cite{o2017variability, pan2019precipitation, mariotti2020windows, nardi2020skillful}. For instance, some climate conditions may impose persistent constraints on seasonal variability, offering opportunities of high predictability; while some other conditions may amplify the state estimation uncertainty, and introduce \textit{forecast busts} \cite{rodwell2013characteristics}.
Retrospective forecasts employing limited ensemble members often fail to give conclusive answer to this predictability-state dependency relationship.
Here, we examine if the CVAE model can 
leverage the rich information from climate simulations to shed light on this problem.
The results in turn verify if we make reasonable forecast uncertainty estimations using the CVAE methodology. 



\begin{figure}[hbt!]
\centering
\includegraphics[width=\linewidth]{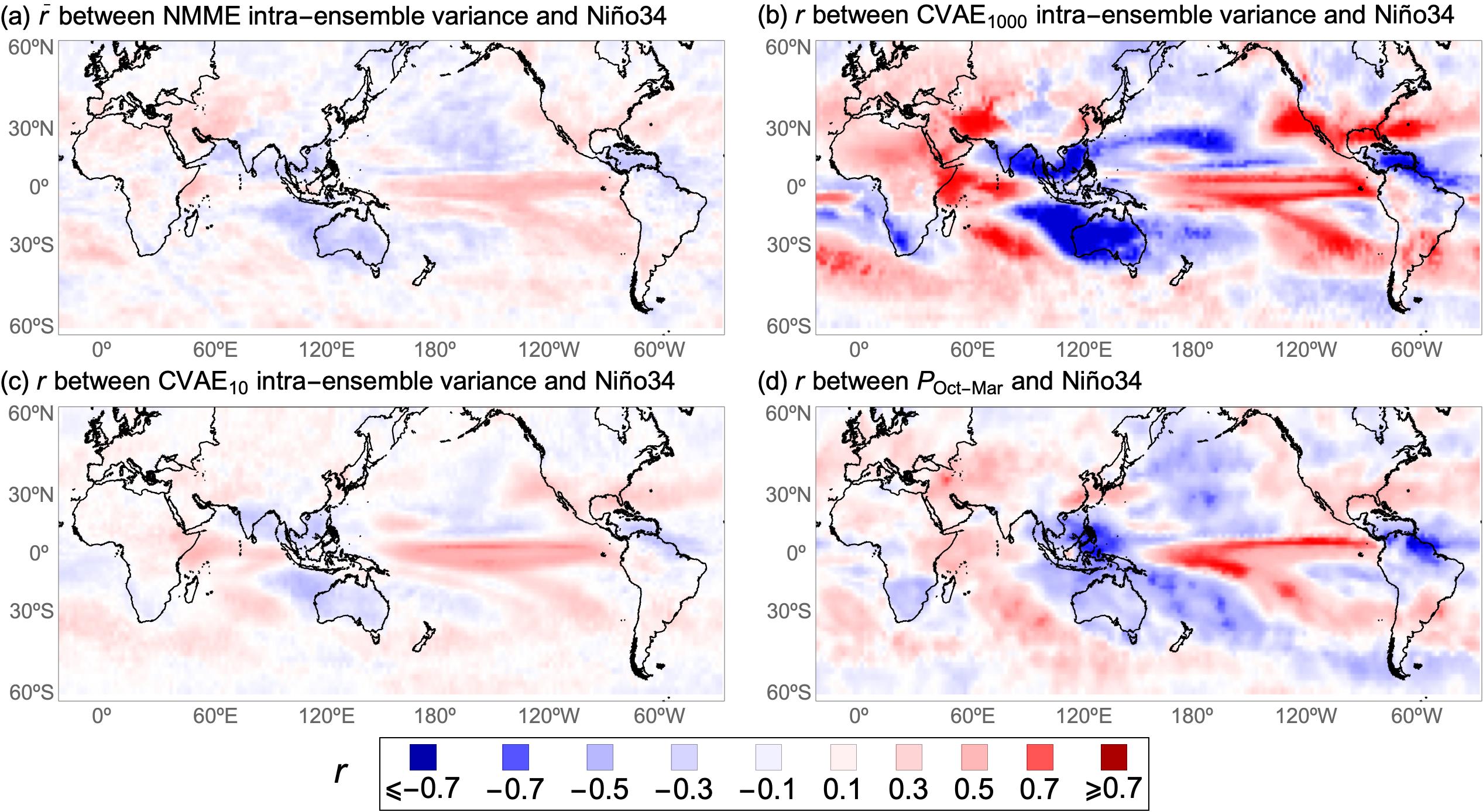}
\caption{Correlating the October-March precipitation forecast intra-ensemble variance or precipitation observation with forecast starting time (July) Ni\~{n}o3.4 index.
(a), correlation coefficient ($r$) between NMME dynamical forecast intra-ensemble variance and the Ni\~{n}o3.4 index. Results are calculated for each of the four NMME models, the mean values are plotted. 
(b), $r$ between CVAE forecast intra-ensemble variance and the Ni\~{n}o3.4 index. The CVAE forecast applies 1000 ensemble members.
(c), similar as (b) but for 10-ensemble member CVAE forecast. 
(d), $r$ between October-March mean precipitation observation and Ni\~{n}o3.4 index of previous July. 
All the correlations are calculated at 2$^{\circ}\times2^{\circ}$ grid scale for the period of 1982 to 2017.
}
\label{fig7: ninop}
\end{figure}

The analysis here focuses on the dominant seasonal variability mode of ENSO and its impact on precipitation predictability.  
To infer this impact, 
we apply the forecast intra-ensemble variance
as an indicator of predictability. 
We correlate this intra-ensemble variance with the forecast starting time ENSO status (measured using the Ni\~{n}o3.4 index) for the period of 1982 to 2017.  Results are shown in Fig.~\ref{fig7: ninop}.

The intra-ensemble variance reflects the growth of state estimation uncertainty in model state space.
In NMME forecast, we explicitly compute this growth of state estimation uncertainty, starting from an ensemble of initial state estimates that are loosely constrained by observations. In CVAE forecast, we adopt a top-down strategy, that is, we discard explicit modeling of the processes that are less predictable or less relevant in seasonal forecasts, and represent their impacts using variational inference. 
Given their distinct representation of uncertainty, it is quite surprising that, despite the magnitude of difference, the NMME and the CVAE model show very similar correlation patterns between Ni\~{n}o3.4 and their forecast intra-ensemble variances (Fig. ~\ref{fig7: ninop}a, b).

The relatively weak correlation between the NMME intra-ensemble variance and the Ni\~{n}o3.4 index (Fig.~\ref{fig7: ninop}a) may reflect the ensemble size limitation in dynamical forecast.
Based on limited ensemble members, the intra-ensemble variance estimation can be seriously influenced by sampling variability noise. 
To support this argument, we draw analog from the CVAE forecast: as we decrease the ensemble size of CVAE forecast from 1,000 to 10, 
the resulting correlation map (Fig.~\ref{fig7: ninop}c) shows similar level of correlation magnitude as in Fig.~\ref{fig7: ninop}a.

The discussion above suggests the necessity for applying large ensembles to  infer the predictability limit and its connections with 
climate variability signals. While this practice is prohibitively costly in dynamical forecast, the CVAE model offers useful analogical insights.
Compared to the NMME result, the correlation map between the CVAE intra-ensemble variance and the Ni\~{n}o3.4 index  gives a more definitive answer to how ENSO modulates predictability.
In Fig.~\ref{fig7: ninop}b, for the negative correlated regions, such as Southeast Asia, Australia, and South Africa, forecasts starting in El Ni\~{n}o Julys tend to give a narrower ensemble spread in predicting the mean boreal winter precipitation, which offer an opportunity of predictability;
contrarily,
predictions starting in La Ni\~{n}a Julys tend to give a wider forecast spread, highlighting the  
necessity of having larger ensembles to sample the less-predictable climate.
The opposite analysis applies for the positive correlated regions: El Ni\~{n}o July forecasts tend to give a wider ensemble spread and less predictability, while La Ni\~{n}a July forecasts tend to give a narrower ensemble spread and more predictability. 

Finally, we notice a strong resemblance between correlating Ni\~{n}o3.4 with intra-ensemble variance (Fig.~\ref{fig7: ninop}a, b, c) and correlating Ni\~{n}o34 with mean precipitation amount (Fig.~\ref{fig7: ninop}d). This suggests a global pattern for the considered seasonal forecast setting: The abnormally high precipitation associated with ENSO is accompanied by a larger forecast uncertainty, while abnormally low precipitation associated with ENSO is accompanied by a smaller forecast uncertainty.

\section{Concluding Remarks}
\label{Section6}


We have developed a deep generative model that draws on a wealth of existing climate simulations for probabilistic seasonal forecasts. The model offers a competitive quasi-global scale seasonal forecast benchmark, and a powerful tool for disentangling the dynamical forecast limitations imposed by initialization errors, model formulation errors, and internal climate variability.
Three of the key contributions here are summarized as follows:
\begin{enumerate}
    \item We successfully extend the data-driven seasonal forecasting paradigm to quasi-global scale. 
    \item We demonstrate that probabilistic machine learning, in particular, deep learning-based variational inference, is a powerful tool for leveraging the rich information from climate simulations to inform seasonal forecast and forecast uncertainty.  
    \item We provide efficient approaches for verifying and diagnosing the ever-complicated dynamical forecast systems, pinpointing 
    clear paths toward forecast improvement. 
\end{enumerate}
This work is enabled by  recent advances in variational-based deep learning models, and the consistent
development and maintenance of climate simulations by the climate community. 
Exploring novel learning paradigms that explicitly learn from each GCM for better ensembles, and verifying advanced neural network parametric forms that better suit spherical climate signals are left for future work.




\section*{Acknowledgments}
This work was performed under the auspices of the U.S. Department of Energy by Lawrence Livermore National Laboratory under contract DE-AC52-07NA27344. Lawrence Livermore National Security, LLC. The views expressed here do not necessarily reflect the opinion of the United States Government, the United States Department of Energy, or Lawrence Livermore  National Laboratory. This work was supported by LLNL Laboratory Directed Research and Development project 19-ER-032.
This document is released with IM tracking number
LLNL-JRNL-810385.

\newpage
\section*{Appendix}


\section*{Derivation of the evidence lower bound equation}
\label{ELBO Derivation}
Equation \ref{ELBO_derivation} gives a step-by-step derivation for the evidence lower bound equation.  Compared to a commonly-used strategy of starting from the viewpoint of minimizing the Kullback-Leibler divergence between posterior $P(Z|X,Y)$ and the variational distribution families, the derivation here focuses on decomposing the conditional marginal log likelihood $\log p_{\theta}(Y|X)$ into the ELBO and the residual term. As the result suggests, by maximizing the ELBO, we will approximately maximize $\log p_{\theta}(Y|X)$ while minimizing $KL\big(q_{\phi}(Z|X,Y)\Vert P(Z|X,Y)\big)$.

\begin{equation}
\begin{split}
&\log p_{\theta}(Y|X)=\int\displaylimits_{Z\sim q_{\phi}}   \log \bigg(\overbrace{\frac{p_{\psi}(Y|X,Z)p(Z|X)}{P(Z|X,Y)}}^{\text{Bayes Rule}}
\cdot
\overbrace{\frac{q_{\phi}(Z|X,Y)}{q_{\phi}(Z|X,Y)}}^{\text{Constant 1}}  \bigg)q_{\phi}(Z|X,Y)dZ\\
&=\int\displaylimits_{Z\sim q_{\phi}} \log p_{\psi}(Y|X,Z)q_{\phi}(Z|X,Y)dZ-
\int\displaylimits_{Z\sim q_{\phi}} \log \frac{q_{\phi}(Z|X,Y)}{p(Z|X)}
q_{\phi}(Z|X,Y)dZ+
\int\displaylimits_{Z\sim q_{\phi}} \log \frac{q_{\phi}(Z|X,Y)}{P(Z|X,Y)}
q_{\phi}(Z|X,Y)dZ\\
&=
\underbrace{\underbrace{E_{Z\sim q_{\phi}} [\log p_{\psi}(Y|X,Z)]}_{\text{Reconstruction term}}-
	\underbrace{KL\big(q_{\phi}(Z|X,Y)
		\Vert
		p(Z|X)\big)}_{\text{Regularization term}}}_{\text{Evidence lower bound}}+
\underbrace{KL\big(q_{\phi}(Z|X,Y)
	\Vert
	P(Z|X,Y)\big)}_{\text{Residual term}}\\
\end{split}
\label{ELBO_derivation}
\end{equation}

\section*{Climate simulation data}

Table~A1 and A2 show the climate simulation data sources from the  Phase 5 \cite{taylor2012overview} and Phase 6 \cite{eyring2016overview} of the Coupled Model Intercomparison Project. 
\begin{table}
\centering
\caption{Sample sources from the Phase 5 of the Coupled Model Intercomparison Project}
\label{tab:cmip5}
\resizebox{!}{0.3\textwidth}{
\begin{threeparttable}
\begin{tabular}{ccccccccccc}
\hline
\multirow{2}{*}{GCM}&\multicolumn{4}{c}{DECK$^\dagger$ and historical}&&\multicolumn{4}{c}{Future scenarios}&\multirow{2}{*}{Total}\\
\cline{2-5}
\cline{7-10}
&\text{1pctCO$_{2}$$^{\dagger\dagger}$} & \text{abrupt4$\times$CO$_{2}$${^\ddagger}$} & \text{historical} & \text{piControl${^{\ddagger\ddagger}}$} &&
   \text{rcp26$^{\mathsection}$} & \text{rcp45$^{\mathsection\mathsection}$} & \text{rcp60$^{*}$} & \text{rcp85$^{**}$} & 
\\
\hline
\text{ACCESS1.0} & 140 & 150 & 156 & 500 &   & 0 & 95 & 0 & 95 & 1136 \\
 \text{ACCESS1.3} & 140 & 151 & 468 & 500 &   & 0 & 95 & 0 & 95 & 1449 \\
 \text{BCC-CSM1.1} & 140 & 150 & 489 & 500 &   & 295 & 295 & 0 & 295 & 2164 \\
 \text{BCC-CSM1.1-M} & 140 & 150 & 489 & 400 &   & 95 & 95 & 95 & 95 & 1559 \\
 \text{CanESM2} & 140 & 150 & 312 & 0 &   & 95 & 0 & 0 & 0 & 697 \\
 \text{CMCC-CESM} & 0 & 0 & 156 & 277 &   & 0 & 0 & 0 & 0 & 433 \\
 \text{CNRM-CM5} & 140 & 150 & 1404 & 850 &   & 95 & 295 & 0 & 675 & 3609 \\
 \text{CNRM-CM5-2} & 490 & 140 & 156 & 350 &   & 0 & 0 & 0 & 0 & 1136 \\
 \text{CSIRO-Mk3.6.0.} & 140 & 150 & 1560 & 500 &   & 950 & 1550 & 950 & 1550 & 7350 \\
 \text{GFDL-CM2-P1} & 0 & 0 & 0 & 0 &   & 0 & 350 & 0 & 0 & 350 \\
 \text{GFDL-CM3} & 140 & 150 & 438 & 0 &   & 95 & 0 & 0 & 95 & 918 \\
 \text{GFDL-ESM2G} & 500 & 300 & 145 & 500 &   & 95 & 95 & 95 & 0 & 1730 \\
 \text{GFDL-ESM2M} & 500 & 0 & 145 & 500 &   & 0 & 0 & 95 & 95 & 1335 \\
 \text{GISS-E2-H} & 453 & 453 & 2340 & 1842 &   & 885 & 4519 & 285 & 1075 & 11852 \\
 \text{GISS-E2-H-CC} & 0 & 0 & 161 & 251 &   & 0 & 95 & 0 & 95 & 602 \\
 \text{GISS-E2-R} & 453 & 302 & 2178 & 1794 &   & 885 & 4615 & 285 & 590 & 11102 \\
 \text{GISS-E2-R-CC} & 0 & 0 & 161 & 251 &   & 0 & 95 & 0 & 95 & 602 \\
 \text{MIROC4h} & 0 & 0 & 168 & 0 &   & 0 & 90 & 0 & 0 & 258 \\
 \text{MIROC5} & 140 & 151 & 815 & 0 &   & 250 & 345 & 345 & 345 & 2391 \\
 \text{MIROC-ESM} & 140 & 150 & 468 & 630 &   & 95 & 295 & 95 & 95 & 1968 \\
 \text{MIROC-ESM-CHEM} & 0 & 0 & 156 & 255 &   & 95 & 95 & 95 & 95 & 791 \\
 \text{MPI-ESM-LR} & 150 & 150 & 468 & 1000 &   & 485 & 190 & 0 & 485 & 2928 \\
 \text{MPI-ESM-MR} & 150 & 150 & 468 & 1000 &   & 95 & 285 & 0 & 95 & 2243 \\
 \text{MPI-ESM-P} & 140 & 150 & 156 & 1156 &   & 0 & 0 & 0 & 0 & 1602 \\
 \text{MRI-CGCM3} & 140 & 150 & 780 & 500 &   & 95 & 95 & 95 & 95 & 1950 \\
 \hline
 \text{Total} & 4376 & 3297 & 14237 & 13556 &   & 4605 & 13589 & 2435 & 6060 & 62155
   \\
 \hline
\end{tabular}
\begin{tablenotes}
            \item[$\dagger$] DECK is abbreviation for the Diagnostic, Evaluation and Characterization of Klima experiments. DECK and historical simulation are common
            experiments across different phases of CMIP.\\
            \item[$\dagger\dagger$] A simulation forced by a 1\%yr$^{-1}$ CO$_2$ increase.\\
            \item[$\ddagger$] A simulation forced by an abrupt quadrupling of CO$_2$.\\ 
            \item[$\ddagger\ddagger$] Pre-industrial control simulation.\\
            \item[$\mathsection$, $\mathsection\mathsection$, $*$, $**$] Future simulation following a representative concentration pathway (rcp) that radiative forcing reach peak at about 2.6 W/m${^2}$/4.5 W/m${^2}$/6 W/m${^2}$/8.5 W/m${^2}$ before 2100 and decline. 
\end{tablenotes}
\end{threeparttable}
}
\end{table}

\begin{table}
\centering
\caption{Sample sources from the Phase 6 of the Coupled Model Intercomparison Project}
\label{tab:cmip6}
\resizebox{!}{0.2\textwidth}{
\begin{threeparttable}
    \begin{tabular}{ccccccccccccccccc}
\hline
\multirow{2}{*}{GCM}&\multicolumn{4}{c}{DECK$^\dagger$ and historical}&&\multicolumn{7}{c}{Future scenarios}&\multirow{2}{*}{Total}\\
\cline{2-5}
\cline{7-13}
&\text{1pctCO$_2$$^\dagger$} & \text{abrupt4$\times$CO$_2$$^\dagger$} & \text{historical} & \text{piControl$^\dagger$} &&
   \text{ssp1-1.9$^{\dagger\dagger}$} &
   \text{ssp1-2.6$^{\ddagger}$}& \text{ssp2-4.5$^{\ddagger\ddagger}$}& \text{ssp3-7.0$^{\mathsection}$} & \text{ssp4-3.4$^{\mathsection\mathsection}$} & \text{ssp4-6.0$^{*}$} &
   \text{ssp5-8.5$^{**}$} \\
\hline
 \text{ACCESS-CM2\tnote{i}} & 150 & 150 & 330 & 500 &   & 0 & 86 & 86 & 86 & 0 & 0 & 86 & 1474
   \\
 \text{ACCESS-ESM1-5} & 150 & 150 & 330 & 0 &   & 0 & 258 & 172 & 258 & 0 & 0 & 258 &
   1576 \\
 \text{BCC-CSM2-MR} & 0 & 0 & 495 & 600 &   & 0 & 86 & 86 & 86 & 0 & 0 & 86 & 1439 \\
 \text{BCC-ESM1} & 0 & 0 & 495 & 0 &   & 0 & 0 & 0 & 0 & 0 & 0 & 0 & 495 \\
 \text{CanESM5} & 151 & 151 & 4125 & 0 &   & 430 & 774 & 860 & 860 & 430 & 430 & 774
   & 8985 \\
 \text{CESM2} & 150 & 999 & 1485 & 1200 &   & 0 & 86 & 86 & 172 & 0 & 0 & 172 & 4350
   \\
 \text{CESM2-WACCM} & 150 & 150 & 495 & 499 &   & 0 & 86 & 86 & 168 & 0 & 0 & 86 &
   1720 \\
 \text{CNRM-CM6-1} & 150 & 150 & 3795 & 500 &   & 0 & 516 & 516 & 516 & 0 & 0 & 516 &
   6659 \\
 \text{CNRM-CM6-1-HR} & 150 & 0 & 165 & 300 &   & 0 & 0 & 0 & 0 & 0 & 0 & 0 & 615 \\
 \text{CNRM-ESM2-1} & 600 & 450 & 825 & 500 &   & 430 & 172 & 258 & 172 & 258 & 258 &
   258 & 4181 \\
 \text{GFDL-CM4} & 150 & 0 & 0 & 500 &   & 0 & 0 & 86 & 0 & 0 & 0 & 86 & 822 \\
 \text{GFDL-ESM4} & 0 & 150 & 165 & 500 &   & 86 & 86 & 86 & 86 & 0 & 0 & 86 & 1245 \\
 \text{GISS-E2-1-G} & 453 & 553 & 3795 & 2027 &   & 0 & 0 & 0 & 0 & 0 & 0 & 0 & 6828 \\
 \text{GISS-E2-1-G-CC} & 0 & 0 & 165 & 165 &   & 0 & 0 & 0 & 0 & 0 & 0 & 0 & 330 \\
 \text{GISS-E2-1-H} & 151 & 302 & 2475 & 401 &   & 0 & 0 & 0 & 0 & 0 & 0 & 0 & 3329 \\
 \text{GISS-E2-2-G} & 151 & 151 & 0 & 151 &   & 0 & 0 & 0 & 0 & 0 & 0 & 0 & 453 \\
 \text{IPSL-CM6A-LR} & 150 & 900 & 5280 & 1450 &   & 516 & 716 & 774 & 946 & 172 & 602
   & 716 & 12222 \\
 \text{MIROC6} & 150 & 250 & 1650 & 800 &   & 86 & 258 & 258 & 258 & 86 & 86 & 258 &
   4140 \\
 \text{MIROC-ES2L} & 150 & 150 & 495 & 500 &   & 86 & 86 & 86 & 86 & 0 & 0 & 86 & 1725
   \\
 \text{MPI-ESM1-2-HR} & 0 & 0 & 0 & 0 &   & 0 & 172 & 86 & 860 & 0 & 0 & 172 & 1290 \\
 \text{MRI-ESM2-0} & 0 & 302 & 0 & 0 &   & 86 & 0 & 86 & 344 & 0 & 0 & 286 & 1104 \\
   \hline
 \text{Total} & 3006& 4958& 26565& 10593&& 1720& 3382& 3612& 4898& 946& 1376& 3926& 64982\\
 \hline
\end{tabular}
\begin{tablenotes}
            \item[$\dagger$] Same as Table~A1.\\ 
            \item[$\dagger\dagger$, $\ddagger$, $\ddagger\ddagger$, $\mathsection$, $\mathsection\mathsection$, 
            $*$, $**$] A simulation following a shared socioeconomic pathway (ssp) with target forcing level at  1.9 W/m${^2}$/2.6 W/m${^2}$/4.5 W/m${^2}$/7.0 W/m${^2}$/3.4 W/m${^2}$/6.0 W/m${^2}$/8.5W/m${^2}$.\\
\end{tablenotes}
\end{threeparttable}
}
\end{table}

\section{ROC curve and Area under the ROC curve score}
To construct the ROC curve,
we consider probability forecasts for the categorical events that the  predictand exceeds pre-defined thresholds. The hate rate and false alarm rate are estimated for these events. For instance, consider the event that the predictand is larger than 20\% quantile of the observations or forecasts, the hit rate (HR) and false alarm rate (FAR) are calculated as follows:
\begin{equation}
    \text{HR}_{20\%}=\frac{n\Big((y_{\text{simu}}>y_{\text{simu}}^{20\%})\land (y_{\text{obser}}>y_{\text{obser}}^{20\%})\Big)}{n(y_{\text{obser}}>y_{\text{obser}}^{20\%})}
\end{equation}
\begin{equation}
    \text{FAR}_{20\%}=\frac{n\Big((y_{\text{simu}}>y_{\text{simu}}^{20\%})\land (y_{\text{obser}}\leq y_{\text{obser}}^{20\%})\Big)}{n(y_{\text{obser}}\leq y_{\text{obser}}^{20\%})}
\end{equation}
Here $\text{HR}_{20\%}$ and $\text{FAR}_{20\%}$ are hit rate and false alarm rate for the event that the predictand $Y$ exceeds the 20\% quantile,  
$n(\cdot)$ denotes the number of events, $y_{\text{simu}}>y_{\text{simu}}^{20\%}$ represents the event that the prediction is larger than the 20\% quantile of all the prediction records. $y_{\text{obser}}>y_{\text{obser}}^{20\%}$/$y_{\text{obser}}\leq y_{\text{obser}}^{20\%}$ represents the event that the observation is larger/no larger than the 20\% quantile of the observation records.

In Fig.~A1, we show how to construct the ROC curve and calculate the AUC score using example of winter precipitation forecast for the Southeastern U.S. (23$^{\circ}$N-27$^{\circ}$N, 81$^{\circ}$W-97$^{\circ}$W). For each of the considered model, we calculate the $(\text{HR}, \text{FAR})$ for the events that predictand exceeds the 20\%, 40\%, 60\%, 80\% quantile of the observations or predictions.
We join these points to obtain the ROC curve. 
The area under the ROC curve is thereafter applied to summarize how reliable the model is at predicting these categorical events.

\begin{figure}[hbt!]
    \centering
    \includegraphics[width=1\linewidth]{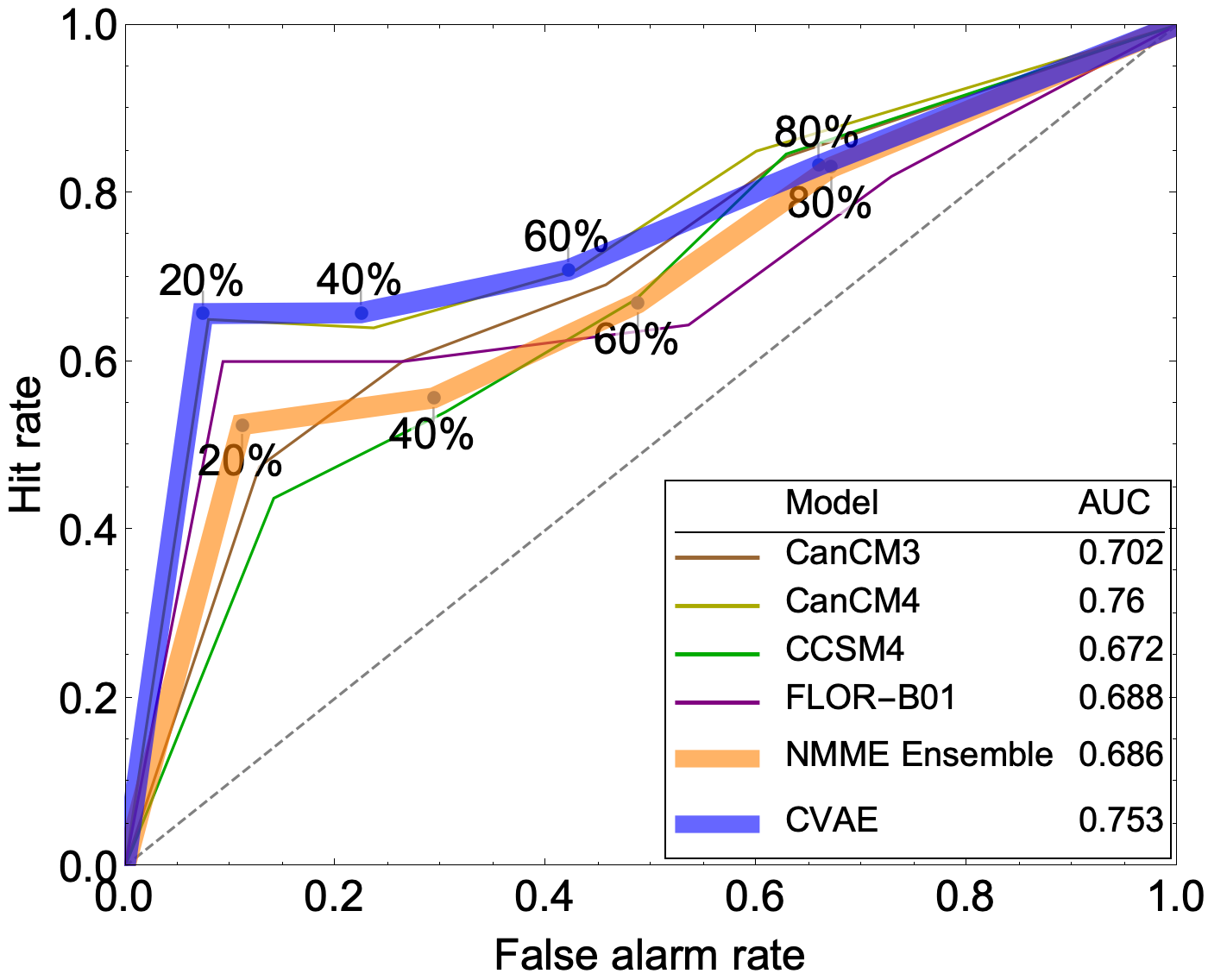}
    \caption{Illustration of the receiver operating characteristic (ROC) curve and the area under the ROC curve score, using example of winter precipitation prediction for the Southeastern U.S. For each of the considered forecasting model (CanCM3, CanCM4, CCSM4, FLOR-B01, their ensemble, and the CVAE forecast), the $(\text{HR}, \text{FAR})$ for the events that predictand exceeds the 20\%, 40\%, 60\%, 80\% quantile of the observations or predictions is calculated. The four  points are joined to form the ROC curve. 
    The area under the ROC curve (AUC) is calculated and labeled on the bottom left.}
    \label{fig:ROCILLUSTRATION}
\end{figure}

\section*{Seasonal prediction skill evaluation using supplementary skill metrics}

Figure~A2 and A3 compare the NMME and CVAE forecast performance  using supplementary deterministic and probabilistic skill metrics of normalized root mean square error (NRMSE) and continuous ranked probability skill score (CRPSS). 

\begin{figure}[hbt!]
\centering
\includegraphics[width=1\linewidth]{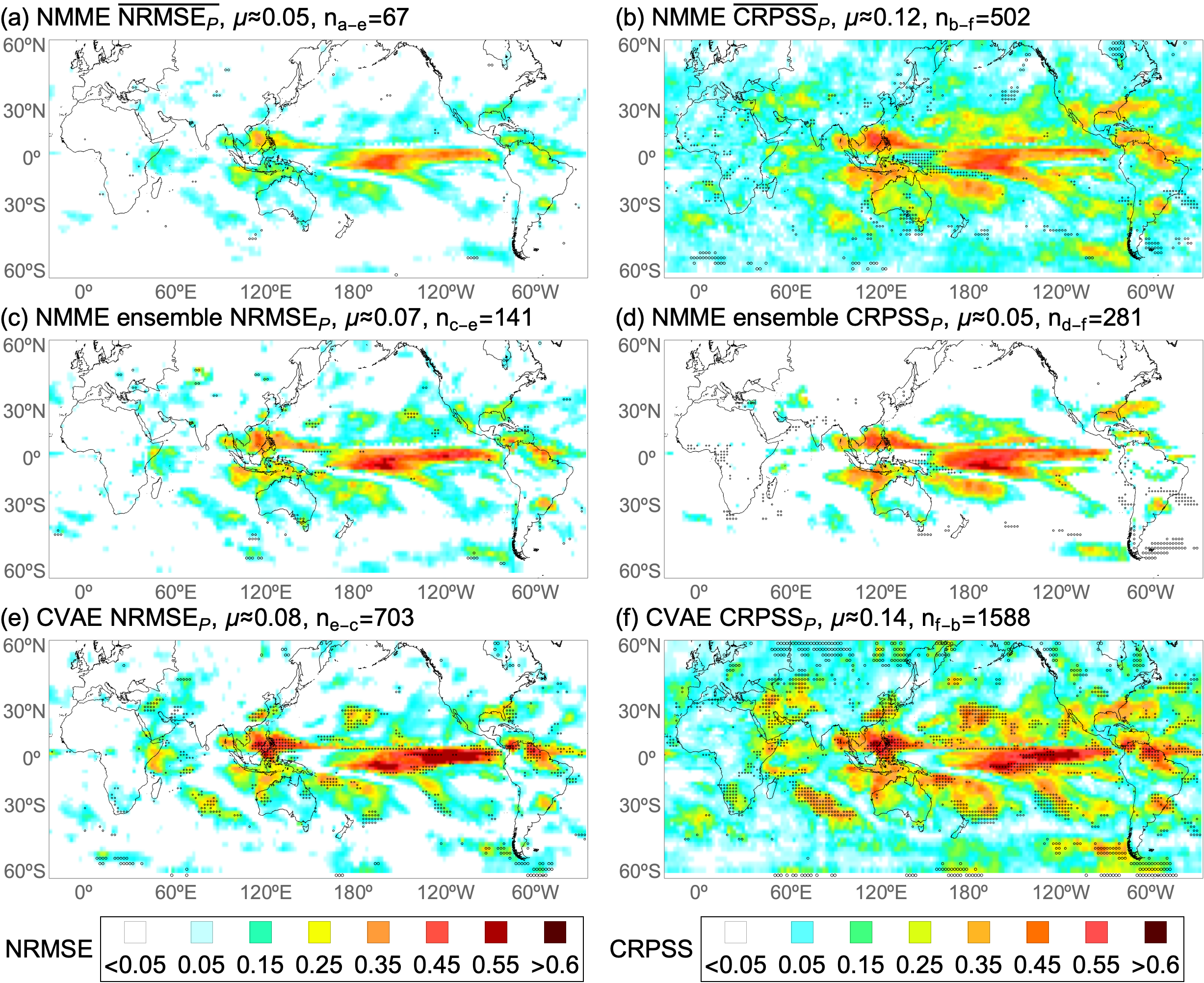}
\caption{Forecast skill for October-March mean precipitation using models starting in July for 1982 to 2018 at $2^{\circ}\times2^{\circ}$ grid scale, measured using supplementary skill metrics of normalized root mean squre error score (NRMSE) and continuous ranked probabilistic skill score (CRPSS).
(a, b), the intra-model mean NRMSE and CRPSS skill score of four NMME models.
(c, d), the NRMSE and CRPSS skill score of the NMME four-model ensemble.
(e, f), the NRMSE and CRPSS skill score of the CVAE model.
Stippling in (a-d) shows locations where the NMME forecasts show significant (95\% confidence level, same for the rest tests) higher skill than CVAE forecasts. 
Stippling in (e) shows locations where the CVAE forecasts show significant higher $r$ score than NMME ensemble forecasts. 
Stippling in (f) shows locations where the CVAE forecasts show significant higher AUC score compared to the mean AUC score of four NMME models. 
The spatial average skill score and the number of stippling are denoted by $\mu$ and $n$ in each sub-figure.}
\label{fig: rmse}
\end{figure}

\begin{figure}[hbt!]
\centering
\includegraphics[width=1\linewidth]{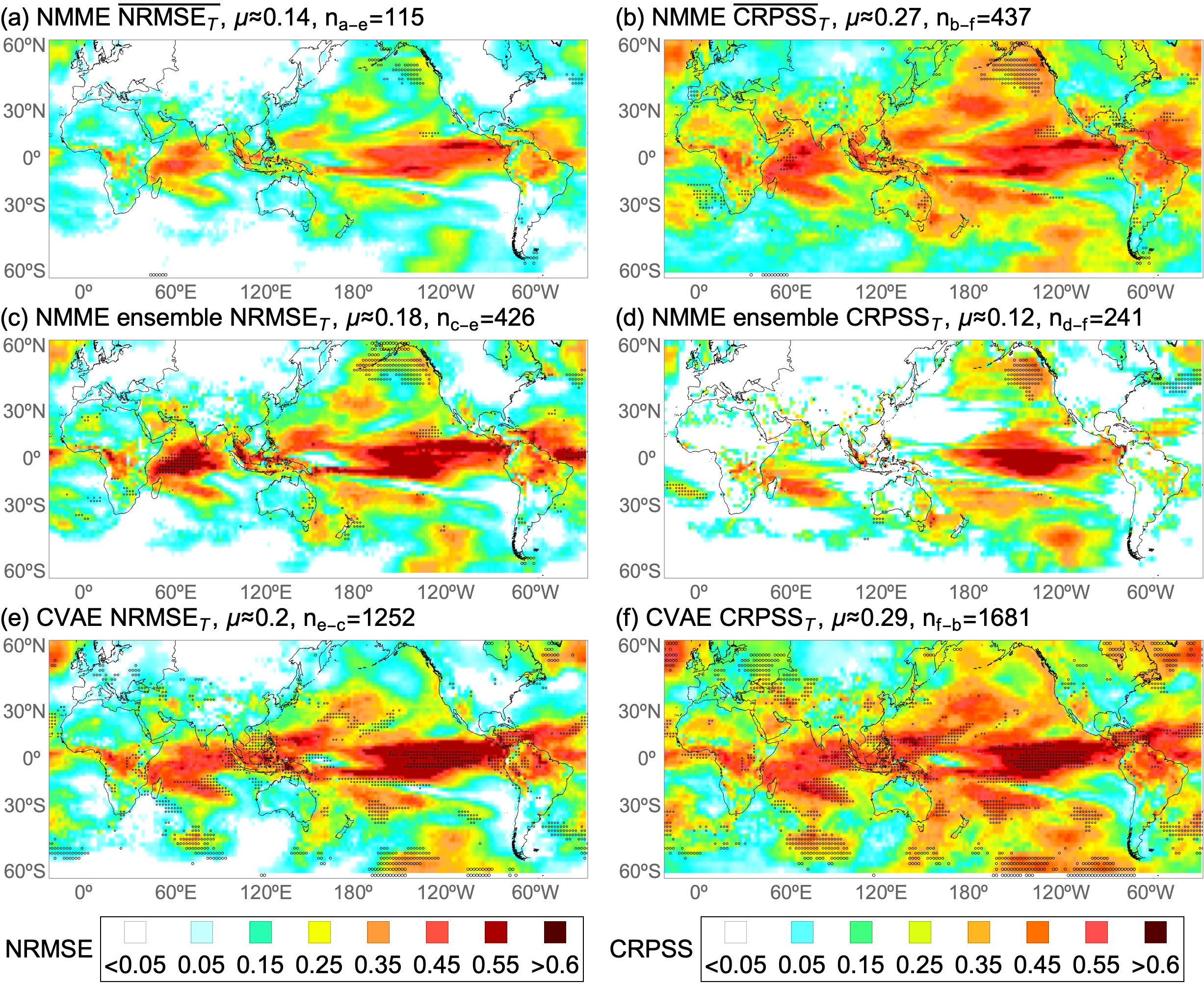}
\caption{Similar as Fig.~A2 but for 2m air temperature forecast.}
\label{fig: supplementaryregion}
\end{figure}

\bibliographystyle{unsrt}
\bibliography{references}

\begin{thebibliography}{10}

\bibitem{bauer2015quiet}
Peter Bauer, Alan Thorpe, and Gilbert Brunet.
\newblock The quiet revolution of numerical weather prediction.
\newblock {\em Nature}, 525(7567):47, 2015.

\bibitem{zhang2019predictability}
Fuqing Zhang, Y~Qiang Sun, Linus Magnusson, Roberto Buizza, Shian-Jiann Lin,
  Jan-Huey Chen, and Kerry Emanuel.
\newblock What is the predictability limit of midlatitude weather?
\newblock {\em Journal of the Atmospheric Sciences}, 76(4):1077--1091, 2019.

\bibitem{palmer2002economic}
Tim~N Palmer.
\newblock The economic value of ensemble forecasts as a tool for risk
  assessment: From days to decades.
\newblock {\em Quarterly Journal of the Royal Meteorological Society: A journal
  of the atmospheric sciences, applied meteorology and physical oceanography},
  128(581):747--774, 2002.

\bibitem{balmaseda2009impact}
M~Balmaseda and D~Anderson.
\newblock Impact of initialization strategies and observations on seasonal
  forecast skill.
\newblock {\em Geophysical research letters}, 36(1), 2009.

\bibitem{mulholland2015origin}
David~P Mulholland, Patrick Laloyaux, Keith Haines, and Magdalena~Alonso
  Balmaseda.
\newblock Origin and impact of initialization shocks in coupled
  atmosphere--ocean forecasts.
\newblock {\em Monthly Weather Review}, 143(11):4631--4644, 2015.

\bibitem{smith2012current}
Doug~M Smith, Adam~A Scaife, and Ben~P Kirtman.
\newblock What is the current state of scientific knowledge with regard to
  seasonal and decadal forecasting?
\newblock {\em Environmental Research Letters}, 7(1):015602, 2012.

\bibitem{pan2019precipitation}
Baoxiang Pan, Kuolin Hsu, Amir AghaKouchak, Soroosh Sorooshian, and Wayne
  Higgins.
\newblock {Precipitation prediction skill for the West Coast United States:
  from short to extended range}.
\newblock {\em Journal of Climate}, 32(1):161--182, 2019.

\bibitem{mariotti2020windows}
Annarita Mariotti, Cory Baggett, Elizabeth~A Barnes, Emily Becker, Amy Butler,
  Dan~C Collins, Paul~A Dirmeyer, Laura Ferranti, Nathaniel~C Johnson, Jeanine
  Jones, et~al.
\newblock Windows of opportunity for skillful forecasts subseasonal to seasonal
  and beyond.
\newblock {\em Bulletin of the American Meteorological Society}, (2020), 2020.

\bibitem{best2015plumbing}
Martin~J Best, Gab Abramowitz, HR~Johnson, AJ~Pitman, Gk~Balsamo, Aaron Boone,
  M~Cuntz, B~Decharme, PA~Dirmeyer, J~Dong, et~al.
\newblock The plumbing of land surface models: benchmarking model performance.
\newblock {\em Journal of Hydrometeorology}, 16(3):1425--1442, 2015.

\bibitem{smith2001disentangling}
Leonard~A Smith.
\newblock Disentangling uncertainty and error: On the predictability of
  nonlinear systems.
\newblock In {\em Nonlinear dynamics and statistics}, pages 31--64. Springer,
  2001.

\bibitem{ding2018skillful}
Hui Ding, Matthew Newman, Michael~A Alexander, and Andrew~T Wittenberg.
\newblock Skillful climate forecasts of the tropical indo-pacific ocean using
  model-analogs.
\newblock {\em Journal of Climate}, 31(14):5437--5459, 2018.

\bibitem{ding2019diagnosing}
Hui Ding, Matthew Newman, Michael~A Alexander, and Andrew~T Wittenberg.
\newblock Diagnosing secular variations in retrospective enso seasonal forecast
  skill using cmip5 model-analogs.
\newblock {\em Geophysical Research Letters}, 46(3):1721--1730, 2019.

\bibitem{ham2019deep}
Yoo-Geun Ham, Jeong-Hwan Kim, and Jing-Jia Luo.
\newblock Deep learning for multi-year {ENSO} forecasts.
\newblock {\em Nature}, 573(7775):568--572, 2019.

\bibitem{altman1992introduction}
Naomi~S Altman.
\newblock An introduction to kernel and nearest-neighbor nonparametric
  regression.
\newblock {\em The American Statistician}, 46(3):175--185, 1992.

\bibitem{kingma2013auto}
Diederik~P Kingma and Max Welling.
\newblock Auto-encoding variational bayes.
\newblock {\em arXiv preprint arXiv:1312.6114}, 2013.

\bibitem{rezende2014stochastic}
Danilo~Jimenez Rezende, Shakir Mohamed, and Daan Wierstra.
\newblock Stochastic backpropagation and approximate inference in deep
  generative models.
\newblock {\em arXiv preprint arXiv:1401.4082}, 2014.

\bibitem{Kihyuk2015}
Kihyuk Sohn, Honglak Lee, and Xinchen Yan.
\newblock Learning structured output representation using deep conditional
  generative models.
\newblock In {\em Advances in Neural Information Processing Systems (NIPS)},
  pages 3483--3491. 2015.

\bibitem{ivanov2018variational}
Oleg Ivanov, Michael Figurnov, and Dmitry Vetrov.
\newblock Variational autoencoder with arbitrary conditioning.
\newblock {\em arXiv preprint arXiv:1806.02382}, 2018.

\bibitem{kingma2019introduction}
Diederik~P Kingma and Max Welling.
\newblock An introduction to variational autoencoders.
\newblock {\em arXiv preprint arXiv:1906.02691}, 2019.

\bibitem{switanek2020present}
Matthew~B Switanek, Joseph~J Barsugli, Michael Scheuerer, and Thomas~M Hamill.
\newblock Present and past sea surface temperatures: a recipe for better
  seasonal climate forecasts.
\newblock {\em Weather and Forecasting}, 35(4):1221--1234, 2020.

\bibitem{doersch2016tutorial}
Carl Doersch.
\newblock Tutorial on variational autoencoders.
\newblock {\em arXiv preprint arXiv:1606.05908}, 2016.

\bibitem{robbins1951stochastic}
Herbert Robbins and Sutton Monro.
\newblock A stochastic approximation method.
\newblock {\em The annals of mathematical statistics}, pages 400--407, 1951.

\bibitem{Bottou2010}
L{\'e}on Bottou.
\newblock Large-scale machine learning with stochastic gradient descent.
\newblock In {\em Proceedings of COMPSTAT'2010}, pages 177--186, 2010.

\bibitem{higgins2017beta}
Irina Higgins, Loic Matthey, Arka Pal, Christopher Burgess, Xavier Glorot,
  Matthew Botvinick, Shakir Mohamed, and Alexander Lerchner.
\newblock {beta-VAE: learning basic visual concepts with a constrained
  variational framework.}
\newblock {\em International Conference on Learning Representations}, 2(5):6,
  2017.

\bibitem{lecun2015deep}
Yann LeCun, Yoshua Bengio, and Geoffrey Hinton.
\newblock Deep learning.
\newblock {\em Nature}, 521(7553):436--444, 2015.

\bibitem{lecun1995convolutional}
Yann LeCun, Yoshua Bengio, et~al.
\newblock Convolutional networks for images, speech, and time series.
\newblock {\em The handbook of brain theory and neural networks},
  3361(10):1995, 1995.

\bibitem{ruthotto2019deep}
Lars Ruthotto and Eldad Haber.
\newblock Deep neural networks motivated by partial differential equations.
\newblock {\em Journal of Mathematical Imaging and Vision}, pages 1--13, 2019.

\bibitem{pan2019improving}
Baoxiang Pan, Kuolin Hsu, Amir AghaKouchak, and Soroosh Sorooshian.
\newblock Improving precipitation estimation using convolutional neural
  network.
\newblock {\em Water Resources Research}, 55(3):2301--2321, 2019.

\bibitem{miao2019improving}
Qinghua Miao, Baoxiang Pan, Hao Wang, Kuolin Hsu, and Soroosh Sorooshian.
\newblock Improving monsoon precipitation prediction using combined
  convolutional and long short term memory neural network.
\newblock {\em Water}, 11(5):977, 2019.

\bibitem{pan2019advancing}
Baoxiang Pan.
\newblock {\em Advancing Precipitation Prediction Using a Composite of Models
  and Data}.
\newblock PhD thesis, UC Irvine, 2019.

\bibitem{weyn2019can}
Jonathan~A Weyn, Dale~R Durran, and Rich Caruana.
\newblock Can machines learn to predict weather? using deep learning to predict
  gridded 500-hpa geopotential height from historical weather data.
\newblock {\em Journal of Advances in Modeling Earth Systems},
  11(8):2680--2693, 2019.

\bibitem{hall2001extratropical}
Nicholas~MJ Hall, Jacques Derome, and Hai Lin.
\newblock The extratropical signal generated by a midlatitude sst anomaly. part
  i: Sensitivity at equilibrium.
\newblock {\em Journal of Climate}, 14(9):2035--2053, 2001.

\bibitem{taylor2012overview}
Karl~E Taylor, Ronald~J Stouffer, and Gerald~A Meehl.
\newblock {An overview of CMIP5 and the experiment design}.
\newblock {\em Bulletin of the American Meteorological Society},
  93(4):485--498, 2012.

\bibitem{eyring2016overview}
Veronika Eyring, Sandrine Bony, Gerald~A Meehl, Catherine~A Senior, Bjorn
  Stevens, Ronald~J Stouffer, and Karl~E Taylor.
\newblock {Overview of the Coupled Model Intercomparison Project Phase 6
  (CMIP6) experimental design and organization}.
\newblock {\em Geoscientific Model Development (Online)}, 9(LLNL-JRNL-736881),
  2016.

\bibitem{zuo2017new}
Hao Zuo, Magdalena~A Balmaseda, and Kristian Mogensen.
\newblock The new eddy-permitting orap5 ocean reanalysis: description,
  evaluation and uncertainties in climate signals.
\newblock {\em Climate Dynamics}, 49(3):791--811, 2017.

\bibitem{huffman1997global}
George~J Huffman, Robert~F Adler, Philip Arkin, Alfred Chang, Ralph Ferraro,
  Arnold Gruber, John Janowiak, Alan McNab, Bruno Rudolf, and Udo Schneider.
\newblock {The global precipitation climatology project (GPCP) combined
  precipitation dataset}.
\newblock {\em Bulletin of the American Meteorological Society}, 78(1):5--20,
  1997.

\bibitem{hersbach2020era5}
Hans Hersbach, Bill Bell, Paul Berrisford, Shoji Hirahara, Andr{\'a}s
  Hor{\'a}nyi, Joaqu{\'\i}n Mu{\~n}oz-Sabater, Julien Nicolas, Carole Peubey,
  Raluca Radu, Dinand Schepers, et~al.
\newblock The era5 global reanalysis.
\newblock {\em Quarterly Journal of the Royal Meteorological Society},
  146(730):1999--2049, 2020.

\bibitem{kirtman2014north}
Ben~P Kirtman, Dughong Min, Johnna~M Infanti, James~L Kinter~III, Daniel~A
  Paolino, Qin Zhang, Huug Van Den~Dool, Suranjana Saha, Malaquias~Pena Mendez,
  Emily Becker, et~al.
\newblock {The North American multimodel ensemble: phase-1
  seasonal-to-interannual prediction; phase-2 toward developing intraseasonal
  prediction}.
\newblock {\em Bulletin of the American Meteorological Society},
  95(4):585--601, 2014.

\bibitem{jahn2012late}
Alexandra Jahn, Kara Sterling, Marika~M Holland, Jennifer~E Kay, James~A
  Maslanik, Cecilia~M Bitz, David~A Bailey, Julienne Stroeve, Elizabeth~C
  Hunke, William~H Lipscomb, et~al.
\newblock {Late-twentieth-century simulation of Arctic sea ice and ocean
  properties in the CCSM4}.
\newblock {\em Journal of Climate}, 25(5):1431--1452, 2012.

\bibitem{merryfield2013canadian}
William~J Merryfield, Woo-Sung Lee, George~J Boer, Viatcheslav~V Kharin, John~F
  Scinocca, Gregory~M Flato, RS~Ajayamohan, John~C Fyfe, Youmin Tang, and
  Saroja Polavarapu.
\newblock {The Canadian seasonal to interannual prediction system. Part I:
  Models and initialization}.
\newblock {\em Monthly weather review}, 141(8):2910--2945, 2013.

\bibitem{msadek2014importance}
Rym Msadek, Gabriel~Andres Vecchi, M~Winton, and RG~Gudgel.
\newblock Importance of initial conditions in seasonal predictions of arctic
  sea ice extent.
\newblock {\em Geophysical Research Letters}, 41(14):5208--5215, 2014.

\bibitem{kirtman2009multimodel}
Ben~P Kirtman and Dughong Min.
\newblock Multimodel ensemble {ENSO} prediction with {CCSM} and {CFS}.
\newblock {\em Monthly Weather Review}, 137(9):2908--2930, 2009.

\bibitem{kingma2014adam}
Diederik~P Kingma and Jimmy Ba.
\newblock Adam: A method for stochastic optimization.
\newblock {\em arXiv preprint arXiv:1412.6980}, 2014.

\bibitem{ioffe2015batch}
Sergey Ioffe and Christian Szegedy.
\newblock Batch normalization: Accelerating deep network training by reducing
  internal covariate shift.
\newblock {\em arXiv preprint arXiv:1502.03167}, 2015.

\bibitem{kharin2003roc}
Viatcheslav~V Kharin and Francis~W Zwiers.
\newblock On the roc score of probability forecasts.
\newblock {\em Journal of Climate}, 16(24):4145--4150, 2003.

\bibitem{hersbach2000decomposition}
Hans Hersbach.
\newblock Decomposition of the continuous ranked probability score for ensemble
  prediction systems.
\newblock {\em Weather and Forecasting}, 15(5):559--570, 2000.

\bibitem{TokinagaEtAl2012}
H.~Tokinaga.
\newblock Slowdown of the walker circulation driven by tropical indo-pacific
  warming.
\newblock {\em Nature}, 491:439--443, 2012.

\bibitem{o2017variability}
Christopher~H O'Reilly, James Heatley, Dave MacLeod, Antje Weisheimer, Tim~N
  Palmer, Nathalie Schaller, and Tim Woollings.
\newblock Variability in seasonal forecast skill of northern hemisphere winters
  over the twentieth century.
\newblock {\em Geophysical Research Letters}, 44(11):5729--5738, 2017.

\bibitem{nardi2020skillful}
Kyle~M Nardi, Cory~F Baggett, Elizabeth~A Barnes, Eric~D Maloney, Daniel~S
  Harnos, and Laura~M Ciasto.
\newblock Skillful all-season s2s prediction of us precipitation using the mjo
  and qbo.
\newblock {\em Weather and Forecasting}, 35(5):2179--2198, 2020.

\bibitem{rodwell2013characteristics}
Mark~J Rodwell, Linus Magnusson, Peter Bauer, Peter Bechtold, Massimo Bonavita,
  Carla Cardinali, Michail Diamantakis, Paul Earnshaw, Antonio Garcia-Mendez,
  Lars Isaksen, et~al.
\newblock Characteristics of occasional poor medium-range weather forecasts for
  europe.
\newblock {\em Bulletin of the American Meteorological Society},
  94(9):1393--1405, 2013.

\end{thebibliography}

\end{document}